\documentclass[preprint,aip]{revtex4-1}
\usepackage{amsmath, amsthm, amssymb, amsfonts}
\usepackage{graphicx}
\usepackage{epstopdf}
\graphicspath {{figure/}}
\usepackage{mathtools}
\usepackage{amsmath}
\usepackage{caption}
\usepackage{subcaption}
\usepackage{amssymb}
\usepackage{amsthm}
\begin{document}
\title{Exact Solution of Hartemann-Luhmann Equation of Motion for a Charged Particle interacting with an Intense Electromagnetic Wave/Pulse}
\author{Shivam Kumar Mishra}
\email[]{shivam.mishra@ipr.res.in}
\author{Sudip Sengupta}
\affiliation{Institute for Plasma Research, Bhat, Gandhinagar, India, 382428}
\affiliation{Homi Bhabha National Institute, Training School Complex, Mumbai, 400094, India}
%
\begin{abstract}
We report an exact solution of the Hartemann-Luhmann equation of motion for a charged particle interacting with an intense electromagnetic wave/pulse. It is found that the radiation reaction force has a significant affect on the charged particle dynamics and the particle shows, on average, a net energy gain over a period of time. Further, using a MATHEMATICA based single particle code, the net energy gained by the particle is compared with that obtained using Landau-Lifshitz and Ford-O'connell equation of motion, for different polarizations of the electromagnetic wave. It is found that the average energy gain is independent of both the chosen model equation and polarization of the electromagnetic wave. Our results thus show, that the simpler and hence analytically tractable Hartemann-Luhmann equation of motion ( as compared to Landau-Lifshitz and Ford-O'connell equation of motion ) is adequate for calculations of practical use (for e.g. energy calculation).

\end{abstract}
\pacs{52.27.Lw,52.30.Ex,52.35.We,51.20.+d}


\maketitle

\section{INTRODUCTION}

An accelerating charged particle is associated with electromagnetic fields that radiates energy irreversibly out to infinity 
\cite{Griffiths,jackson}.
In addition to radiated energy, there exists an electromagnetic field energy called the Schott energy which is localized at the particle and can be exchanged with the particle's mechanical energy
\cite{Schott,Ferris}. 
These fields ( radiation field and the Schott field ) carry energy and momentum and affect the dynamics of the charged particle by providing a self-influence called the self-force or radiation reaction (RR) force \cite{Lorentz,Abraham,Dirac1,Dirac2,bhabha, Wheeler,Rohrlich2,Maxwell,Barut,roharlic1}. Radiation reaction force  significantly affects the charged particle dynamics when the power radiated by the particle becomes comparable to the instantaneous rate of change of  energy of the particle \cite{Shen,hadad,noble,Vikrampagi}.
In this scenario, the Lorentz force equation is not an appropriate choice for investigating the  charged particle dynamics. For a complete description of the charged particle dynamics, we need an equation which incorporates the effect of radiation reaction. In this context, a long ongoing debate has suggested a number of equations of motion {\it viz.} Lorentz-Abraham-Dirac(LAD)\cite{Dirac2}, Ford-O'connell\cite{FOC1,FOC2,FOC3}, Eliezer\cite{Eliezer}, Landau-Lifshitz\cite{Landau1}, M.O.Papas\cite{MO}, Caldirola\citep{Caldirola}, Hartemann-Luhmann\citep{Hartemann}, Yaghjian\cite{Yaremko}, Sokolov\cite{Sokolov} et. al. etc. It is worth noting here that except for the LAD equation, the other equations have not been derived from first principles. Thus, the LAD equation is the most basic equation for describing the interaction of a spin less point charged particle with an electromagnetic field and it can be written as
\begin{equation}\label{lad}
\dot{u}^{\alpha} = \frac{e}{mc} F^{\alpha \beta} u_{\beta} + g^{\alpha}
%
\end{equation}
where $F^{{\alpha}{ \beta}}$ is the electromagnetic field tensor, $u^{\alpha} ={\gamma (c,\vec{v})}$ is the four-velocity of the charge, $\gamma = 1 / \sqrt{1-(v/c)^2}$,
 $\tau_{0}=( 2 / 3)(e^2 / m c^3)$ and the dot represents differentiation with respect to proper time 
 $\tau$. 
We use the metric convention $(+1,-1,-1,-1)$. For the LAD equation the radiation reaction term $g^{\alpha}$ is given by,
\begin{equation}\label{rr}
g^{\alpha}=  \tau_{0} \left({\ddot{u}}^{\alpha} + \frac{ \dot{u}^{\beta} \dot{u}_{\beta}}{c^2} {{u}}^{\alpha}\right)
\end{equation}
In the above equation, the first and the second term on the right hand side respectively represent the
radiation reaction force due to the Schott energy and the radiated energy\cite{Schott,rohrlich3}. The presence of the Schott term in the LAD equation results in well known unphysical problems like pre-acceleration and runway solution\cite{Griffiths,jackson,roharlic1,Hartemann2}. Using the condition that
the four acceleration is orthogonal to the four velocity \cite{Landau1}, the radiation reaction term in Eq.(\ref{rr}) can be rewritten as
\begin{equation}\label{rr1}
g^{\alpha}= \tau_{0} \left( \delta_{\beta}^{\alpha}- \frac{ {u}^{\alpha} {u}_{\beta}}{c^2}\right){\ddot{u}}^{\beta}
\end{equation}
From Eq.(\ref{rr1}), it can be seen that the Schott term is of order unity whereas the radiated term is of order $\sim \gamma^{2}$. So in the ultra relativistic case, where radiation reaction becomes important, Schott term may be neglected in comparison with the radiated term. Thus, Eq.(\ref{lad}) can be written as
\begin{equation}\label{hl}
\dot{u}^{\alpha} = \frac{e}{mc} F^{\alpha \beta} u_{\beta} + \tau_{0}\frac{ {\dot{u}}^{\beta} {\dot{u}}_{\beta}}{c^2}{u}^{\alpha}
\end{equation}
Eq.(\ref{hl}) was first derived by Hartemann and Luhmann within the framework of classical electrodynamics and does not suffer from unphysical problems like pre-acceleration and runway solutions\cite{Hartemann}. 

Following Landau-Lifshitz, in the non-relativistic regime, the radiation reaction term may be considered small as compared to the Lorentz force term, provided  
%
the typical wavelength $\lambda$ and the typical field amplitude $F$ of the external electromagnetic field fulfill the following two conditions; 
\begin{equation}\label{con}
\lambda>> \frac{e^{2}}{m c^2}, F<< \frac{m^{2}c^{4}}{e^{3}}
\end{equation}
In order to perform an analogous reduction of degree in the relativistic case, the conditions (\ref{con}) have to be fulfilled in the instantaneous rest frame of the charged particle\cite{Landau1,Bild}. This allows for reduction of degree of the Hartemann-Luhmann equation, by substituting the acceleration terms in the radiation reaction force with the Lorentz force term.  Performing this substitution, Eq.(\ref{hl}) can finally be written as  
\begin{equation}\label{mhl}
\dot{u}^{\alpha} = \frac{e}{mc} F^{\alpha \beta} u_{\beta} - \tau_{0} \frac{e^{2} }{ m^{2} c{^4}} u^{\alpha} F^{ \beta \gamma}F_{\gamma \delta} u^{\delta} u_{\beta}
\end{equation}

Eq.(\ref{mhl}) is the modified form of Hartemann-Luhmannn equation of motion. 
The purpose of this article is to study the charged particle dynamics in the presence of a relativistically intense elliptically polarized electromagnetic wave/pulse using the above new form of Hartemann-Luhmann equation, as given by Eq.(\ref{mhl}). This study gives a clearer understanding of particle dynamics and energy gain during wave particle interaction, than previous studies that have been done by several authors using the Landau-Lifshitz equation of motion \cite{hadad,noble,burton,Keitel_1998,Piazza1,Piazza-prl, R.Ondarza}.
In section \ref{solution} of this article, the new form of Hartemannn-Luhmann equation (hereinafter called modified Hartmann-Luhmann equation) has been solved analytically for energy, momentum, and position of a particle which is interacting with a relativistically intense light wave/pulse. The dynamics of a charged particle interacting with a wave train and a Gaussian laser pulse is  presented in subsections \ref{dynamics_wave} and \ref{dynamics_pulse} respectively. In section \ref{comparison}, the average energy gained by the particle as obtained from the modified Hartemann-Luhmann equation is compared with that obtained by numerically solving the  Landau-Lifshitz and Ford-O'connell equation of motion. Finally in section \ref{summary}, we present a summary of our results.


\section{Solution of Hartemann-Luhmann Equation} \label{solution}
In dimensionless form, the modified Hartemann-Luhmann equations, governing the dynamics of a charged particle interacting with an intense electromagnetic wave are 
\begin{equation} \label{motion}
\frac{d \vec{p}}{d t} = \left( \vec{E} + \vec{\beta} \times \vec{B} \right) - \tau_{0} \gamma^{2} \left\{ \left( \vec{E} + \vec{\beta} \times \vec{B} \right)^{2} - \left( \vec{E}.\vec{\beta} \right)^{2} \right\} \vec{\beta} 
\end{equation}
\begin{equation} \label{energy}
\frac{d \gamma}{d t} = \vec{E}.\vec{\beta} - \tau_{0} \gamma^{2} \left\{ \left( \vec{E} + \vec{\beta} \times \vec{B} \right)^{2} - \left( \vec{E}.\vec{\beta} \right)^{2} \right\}
\end{equation}
where the symbols have their usual meanings and the normalizations used are  $p \rightarrow p/mc$, $E \rightarrow eE/m \omega c$, $B \rightarrow eB/m \omega c$, $\tau_{0} \rightarrow \omega \tau_{0}$ and $\beta = v / c$. The above equations are respectively the spatial and temporal component of Eq.(\ref{mhl}). Consider the interaction of a charged particle with a light wave/pulse which is propagating along the z-direction and defined by the normalized vector potential $\vec{A}(\phi)$ 
($\vec{A} \rightarrow e \vec{A} / m c^{2}$) where $ \phi = t - z + \phi_{0}$ ($t \rightarrow \omega t$ and $z \rightarrow k z$). Here $\omega $ and  $k$ are respectively the frequency and wave number of the light wave, and $\phi_{0}$ is the initial phase of the light wave as seen by the particle. Using $\vec{E} = - \partial \vec{A} / \partial t$ and $\vec{B}= \vec{\nabla} \times \vec{A}$, Eq.(\ref{motion}) and Eq.(\ref{energy}) may be respectively written as 

\begin{equation}\label{motion_1}
\frac{d \vec{p}}{d t} = \left( - \frac{d \vec{A}}{dt}  + \vec{\nabla}(\vec{\beta}.\vec{A}) \right) - \tau_{0} \gamma^{2} \left\{ \left(- \frac{d \vec{A}}{dt}  + \vec{\nabla}(\vec{\beta}.\vec{A}) \right)^{2} - \left( \vec{\beta}.\frac{\partial \vec{A}}{\partial t} \right)^{2} \right\} \vec{\beta}
\end{equation}
\begin{equation}\label{energy_1}
\frac{d \gamma}{d t} = - \beta . \frac{\partial \vec{A}}{\partial t}  - \tau_{0} \gamma^{2} \left\{ \left(- \frac{d \vec{A}}{dt}  + \vec{\nabla}(\vec{\beta}.\vec{A}) \right)^{2} - \left( \vec{\beta}.\frac{\partial \vec{A}}{\partial t} \right)^{2} \right\}
\end{equation}
where $d \vec{A}/ d t = \partial \vec{A}/\partial t + (\beta.\vec{\nabla}) \vec{A}$, and we have used the relation  $\vec{\nabla}(\vec{\beta}.\vec{A}) = (\vec{\beta}.\vec{\nabla}) \vec{A} + \vec{\beta} \times (
 \vec{\nabla} \times \vec{A} )$ as $\beta$ is a function of time alone. Using $\vec{A} = \vec{A}(\phi)$ and transverse nature of the light wave/pulse Eq.(\ref{motion_1}) and Eq.(\ref{energy_1}) may be respectively written as
  
\begin{equation}\label{motion_2}
\frac{\Delta}{\gamma}\vec{p'} = - \left[\vec{A'} \frac{\Delta}{\gamma} + (\vec{\beta}.\vec{A'}) \hat{z}\right] - \tau_{0}\Delta^{2} A'^{2} \vec{\beta}
\end{equation}
\begin{equation}\label{energy_2}
\frac{\Delta}{\gamma}\gamma' = - \beta . \vec{A'} - \tau_{0}\Delta^{2} A'^{2}
\end{equation}
Here prime denotes derivative with respect to $\phi$, $\Delta = \gamma - p_z$ ( $p_z = \gamma \beta_z$ )
and $d/d t = (\Delta/\gamma) d/d \phi$. Subtracting the z-component of equation of motion (Eq.(\ref{motion_2})) from the energy equation (Eq.(\ref{energy_2})) we get 
\begin{equation}\label{delta_prime}
\Delta' = - \tau_{0} \Delta^{2} A'^{2}
\end{equation}
On integrating the above equation and taking $\phi = \phi_0$ at $t = 0$ (and $z=0$) we get 
\begin{equation}\label{delta}
 \Delta = \frac{\Delta_{0}}{1 + \tau_{0} \Delta_{0} I_{1}}
\end{equation}
where $I_{1} = \int_{\phi_{0}}^{\phi}A'^{2} d{\phi}$ and $\Delta_{0} = (\gamma_{0} - p_{z0})$ where 
$\gamma_{0}$ and $p_{z0}$ are respectively the initial energy and z-component of initial momentum of the particle. It is clear from the above equation that in the absence of radiation reaction ({\it i.e.} $\tau_{0} = 0$), $\Delta$ is a constant of motion\cite{Kaw_Kulsrud,Landau1,Gibbon}. Now the perpendicular component of equation of motion (x,y components of Eq.(\ref{motion_2})) gives
\begin{equation}\label{eqperp1}
\Delta \vec{p'}_{\perp} = - \vec{A'} \Delta - \tau_{0} \Delta^{2} A'^{2} \vec{p}_{\perp}
\end{equation}
Multiplying equation Eq.(\ref{delta_prime}) by $\vec{p}_{\perp}$, subtracting from the above equation and dividing by $\Delta^{2}$ we get
%
\begin{equation}\label{eqperp2}
\left(\frac{\vec{p}_{\perp}}{\Delta}\right)' = -\frac{\vec{A'}}{\Delta}
\end{equation}
which on integration yields the perpendicular component of momentum as
\begin{equation}\label{perp}
\vec{p}_{\perp} = \frac{\Delta}{\Delta_0} \left[ \vec{p}_{\perp0} - (\vec{A} -\vec{A}_{0}) - \tau_{0}\Delta_0 \vec{I}_{2} \right]
\end{equation}
Here $\vec{I}_{2} = \int_{\phi_{0}}^{\phi} \vec{A^{'}} I_{1} d \phi$, $\vec{p}_{\perp 0}$ is the initial perpendicular component of momentum and $\vec{A_{0}} = \vec{A}(\phi_0)$ is the vector potential seen by the particle at $t=0$ (and $z=0$). 

To obtain the longitudinal component of momentum, we begin with the z-component of equation of motion (Eq. (\ref{motion_2})) as
\begin{equation}\label{eqpara1}
\Delta \vec{p'}_{z} = - \vec{\beta}.\vec{A'} - \tau_{0}\Delta^{2} A'^{2}
\end{equation}
Following a similar procedure as used for obtaining the perpendicular component of momentum, {\it i.e.} multiplying Eq.(\ref{delta_prime}) 
$p_z$, subtracting from the above equation and dividing by $\Delta^{2}$ we get
\begin{equation}\label{eqara2}
\left(\frac{p_z}{\Delta}\right)' = -\frac{\vec{p}_{\perp}.\vec{A'}}{\Delta^2}
\end{equation}
Substituting the expressions for $\Delta$ and $\vec{p}_{\perp}$ from Eqs. (\ref{delta})  and  (\ref{perp}) respectively and integrating, we arrive at the final expression for longitudinal momentum as 
\begin{equation}\label{para}
p_{z} = \frac{\Delta}{\Delta_0} \left[ p_{z0} - 
\frac{1}{\Delta_0}\left\{ \vec{p}_{\perp 0} - \frac{1}{2}( \vec{A}- \vec{A_0})\right\}.(\vec{A}- \vec{A_0})  -  {\tau_{0}} \left\{{\vec{p}_{\perp 0}} - (\vec{A}- \vec{A_0})\right\}.\vec{I_2} + \frac{1}{2} \Delta_0 \tau_{0}^{2} I_{2}^{2} \right]
\end{equation}
Equations (\ref{perp}), (\ref{para}) along with $\gamma = p_{z} + \Delta$  ( with $\Delta$ given by equation (\ref{delta}) ) respectively give the momentum and energy as a function of phase $\phi$, for a charged particle interacting with an electromagnetic wave/pulse including radiation reaction effects. Eqs. (\ref{perp}) and (\ref{para}) can be integrated further to obtain the particle positions. The expressions for positions of a charged particle interacting with an elliptically polarized light wave train are presented in the Appendix.
%
%
\section{Dynamics of a charged Particle in an Electromagnetic Wave/Pulse} \label{dynamics}
The well known solution of Lorentz force equation ( {\it i.e.} without the radiation reaction term ) shows that the charged particle motion in an electromagnetic wave/pulse results in no net energy gain by the particle from the wave/pulse. In case of interaction with a finite duration pulse, the dynamics only results in a net displacement of the particle from its initial position\cite{Landau1,Gibbon}. It happens due to the well defined phase relationship between the electric field of the electromagnetic wave and the velocity of the particle. In the presence of radiation reaction force the phase relationship between the electric field and the velocity of the particle is disturbed in such a way that as a result the charged particle gains a net amount of energy. This can be seen from the radiation reaction term in the Hartemann-Luhmann equation which acts in a direction opposite to the velocity vector of the particle. In the next two subsections, we respectively describe the motion of a charged particle interacting with a wave train and a pulse. Results are obtained by numerically solving the Hartemann-Luhmann equations using a MATHEMATICA based single particle code and are also compared with the analytical solutions obtained in section \ref{solution}.
%
%
%
\subsection{DYNAMICS OF A CHARGED PARTICLE IN AN ELECTROMAGNETIC WAVE TRAIN} \label{dynamics_wave}
The vector potential representing a wave train is given by
\begin{equation}\label{wave_train}
A(\phi) = a_{0} \left( \delta \cos(\phi) \hat{x} +  g \sqrt{1-\delta^{2}}\sin(\phi) \hat{y} \right)
\end{equation}
The value of $\delta = (0,1)$ and $\delta=1/\sqrt{2}$ correspond to linear and circular polarization respectively. $g=\pm1$ respectively correspond to right and left handedness of polarization. We now present results which clearly exhibit the effect of radiation-reaction on charged particle dynamics moving in an intense electromagnetic wave train for two different polarizations {\it viz.} linear ($\delta = 1$) and circular ($\delta=1/\sqrt{2}$). 
As stated before, the Hartemann-Luhmann equations are numerically integrated with $a_{0}=500$ and $\tau_0 = 1.8 \times 10^{-8}$ ; the particle starts from origin with initial momenta as $p_{\perp 0} = 0$ and $p_{z0} = -10^{3}$ ( initial energy $\gamma = 10^{3}$ ).
%
%

Fig.1 and Fig.2 represent the evolution of longitudinal and transverse momentum, energy and trajectory of the particle in linearly ($\delta = 1$)  and circularly polarized ($\delta=1/\sqrt{2}$) electromagnetic wave respectively. The red and blue curve respectively represent the solution of the Hartemann-Luhmann and Lorentz force equations. The dashed green curve on the red and blue curve represent the analytical solution. The clear mismatch between the red and blue curve show the strong signature of the radiation reaction force. In fig. 1(a), the blue line represents the longitudinal momentum of the particle without radiation-reaction. This shows that the particle is drifting opposite to the direction of propagation of the wave, whereas red line shows that, inclusion of radiation reaction stops the particle within one cycle of electromagnetic wave and the particle is pushed along the direction of propagation of the wave. 
Fig 1(b) represents the transverse momentum with and without radiation reaction, which shows that the net average transverse momentum with radiation reaction is smaller than the transverse momentum without radiation reaction. Fig. 1(c) represents the evolution of energy of the particle with and without radiation reaction. The red curve shows that in the presence of radiation reaction, energy of the particle shows a drastic change, whereas in the absence of the radiation reaction, the average energy of the particle remains constant. Finally Fig. 1(d) represents the trajectory of the particle with ( red ) and without radiation reaction (blue).
%
The results are qualitatively same for circularly polarized light shown in fig.2. The fig. 3 represents the net average energy gained by the particle for different polarization of the electromagnetic wave. The red, blue and green curves respectively correspond to $\delta=1$, $\delta=1/\sqrt{2}$ and 
$\delta=1/4$. The results show that 
the average energy gain is independent of the polarization of the wave. 
%
%
\subsection{DYNAMICS OF A CHARGED PARTICLE IN THE PRESENCE OF ELLIPTICALLY POLARIZED GAUSSIAN LASER PULSE} \label{dynamics_pulse}
In this subsection the effect of the radiation-reaction has been studied for a elliptically polarized Gaussian laser pulse. The vector potential for the laser pulse is given by 
\begin{equation}\label{pulse}
\vec{A}(\phi)  = a_0 e^{{-\frac{(\phi-\xi_{0})^2}{2 n_{0}}}} \left( \delta  cos(\phi) \hat{x} + g \sqrt{1-\delta^{2}} sin(\phi) \hat{y} \right)
\end{equation}
The effect of radiation-reaction on charged particle dynamics moving in intense electromagnetic laser pulse has been studied for linearly ($\delta = 1$) and circularly($\delta=1/\sqrt{2}$) polarized light. The energy gain has been studied  by taking $n_{0}=10$, $\xi_{0}=n_{0} \pi$, $a_{0}=10^{3}$ and $\tau_0 = 1.8 \times 10^{-8}$; the particle initially starts from the origin with $p_{\perp 0}= 0$, and $p_{z 0} = -10^{3}$. 

Fig.4 and Fig.5 respectively represent the evolution of longitudinal and transverse momentum, energy and trajectory of the particle in linearly ($\delta = 1$)  and circularly polarized ($\delta=1/\sqrt{2}$) electromagnetic light pulse wave. As before, the the red and blue curve represent the solution of the Hartemann-Luhmann and Lorentz force equations respectively. The clear mismatch between the red and blue curve show a strong signature of the radiation reaction force. In fig.4(a), the blue line represents the longitudinal momentum of the particle without radiation-reaction. This shows that the particle is drifting opposite to the direction of propagation of the wave, whereas red line shows the inclusion of radiation reaction stops the particle within laser pulse and the particle is pushed along the direction of propagation of the wave. Fig. 4(b) represents the transverse momentum with and without radiation reaction, which shows that there is net gain in transverse momentum in the presence of radiation reaction. 
Fig. 4(c) represents the change in energy of the particle with and without radiation reaction. The red curve shows that in the presence of radiation reaction, the particle gains energy, which can also be seen from the gain in transverse as well as the longitudinal momentum of the particle, after the pulse has passed over the particle. However, in the absence of the radiation reaction, there is no net gain in energy of the particle from the pulse. Finally Fig. 4(d) represents the trajectory of the particle with (red) and without radiation reaction (blue). 
%
%
The results are qualitatively same for the circularly polarized light as shown in fig.5. As the wave crosses over the particle, fig.6 represents the net energy remaining with the particle in the case of different polarization of the electromagnetic wave. The red,  blue,  and green curves correspond to $\delta = 1,\,\, 1/\sqrt{2},\,\, 1/4$ respectively.
%
%
\section{Comparison of Harteman-Luhmann equation of motion with Landau-Lifshitz and Ford-O'connel} \label{comparison}
The Landau-Lifshitz  and Ford-O'connell equation of motion in normalized form can respectively be written as 
\begin{equation}\label{ll}
\dot{u}^{\alpha} = F^{\alpha \beta} u_{\beta} + \tau_{0} \left(F^{\alpha\beta}_{,\gamma} u^{\gamma} u_{\beta} + F^{\alpha \beta}F_{\beta \gamma} u^{\gamma} - u^{\alpha} F^{ \beta \gamma}F_{\gamma \delta} u^{\delta} u_{\beta}\right)
\end{equation}
and 
\begin{equation}\label{foc}
\dot{u}^{\alpha} = F^{\alpha \beta} u_{\beta} + \tau_{0} \Delta_{\beta}^{\alpha} \dot{f}^{\beta}
\end{equation}
The normalizations is the same as that used for Hartemann-Luhmann equation of motion. Here $F^{\alpha \beta}_{, \gamma} = \partial F^{\alpha \beta} / \partial x^{\gamma}$,
$\Delta_{\beta}^{\alpha} = (\delta^{\alpha}_{\beta} - u^{\alpha} u_{\beta} ) $ and ${f}^{\beta}$ is the Lorentz force. We solve the above equations numerically using a MATHEMATICA based single particle code. 

To test the validity of the Hartemann-Luhmann equation of motion, the energy gained by the particle has been compared with the numerical solution of the  Landau-Lifshitz as well as Ford-O'connell equation of motion, for both the cases {\it viz.} for a wave train and for a Gaussian pulse. 
Fig. 7(a) and Fig. 7(b) respectively show the energy gain for a charged particle interacting with a wave train (with $a_0 = 500$ ) and Gaussian pulse ( with $a_0 = 10^{3}$ ). The initial conditions chosen are the same for both the cases; the particle starts from the origin with initial momentum as $p_{\perp 0} = 0$ and $p_{z0} = -10^{3}$. Here red, blue and green curves respectively represent the energy gain obtained from the solution of Hartemann-Luhmann, Landau-Lifshitz and Ford-O'connell equations of motion. 
%
%
Thus comparison of energy gain shows that the results are independent of the model equation; the contribution of the terms eliminated in the Hartemann-Luhmann equation of motion have a negligible influence on the final energy gain.
%
%
%
\section{Summary and conclusions}\label{summary}
Present day lasers can deliver very high intensities, of the order of  $\sim 10^{22}W/cm^2$. In future, intensities could be extended by more than two orders of magnitude\cite{shi}. It is well known that for intensities of the order of $\sim 10^{22} W/cm^2$, for electrons having initial energy in the $\sim GeV$ range, the radiation reaction force can become comparable or even greater than the applied force \cite{noble}. In this scenario, neglecting radiation reaction force from charged particle dynamics is a serious approximation. 

In this article we have presented a simple and  complete picture of particle dynamics in a relativistically intense electromagnetic wave/pulse including the effects of radiation reaction. In this context it has been shown that for ultra-high intensities radiation-reaction force is basically due to the influence of radiated energy, and the contribution of Schott energy is negligibly small. In order to incorporate the effect of radiation reaction into charged particle dynamics for ultra-high laser intensities Hartemann-Luhmann equation has been analytically solved, for particle dynamics in an elliptically polarized electromagnetic wave/pulse. A comparative study between Hartemann-Luhmann equation and Lorentz Force equation shows that at ultra-high laser intensities, radiation reaction force significantly affects the charged particle dynamics. The radiation reaction force disturbs the phase relationship between the velocity and the electric field of the wave in such a way that the particle starts gaining energy.
%
%
A comparison between the net energy gain with different polarization of electromagnetic wave, shows that the net energy gain is independent of polarization. 
Further, using a MATHEMATICA based single particle code, the analytical results of the Hartemann-Luhmann equation of motion has been verified and the energy gain by the particle is compared with the Landau-Lifshitz and Ford-O'connell equation of motion. A good match of energy gain for these three equations shows that the energy gain is independent of chosen model. 
%
%
\appendix
\section{Analytical Expressions for Particle Positions}
The solution of the integrations ${I}_{1}$, $\vec{I}_{2}$ and expressions for the position of the particle wave train are given by
\begin{equation}
I_{1} = \frac{1}{4} \left[ -2\phi + (2\delta^{2} - 1)\left\{ (sin(2 \phi) - (sin(2 \phi_{0}) \right\} + 2 \phi_{0} \right]
\end{equation}
\begin{equation}
\begin{split}
I_{2x} = \frac{1}{24} \delta((15-6\delta^{2}) sin(\phi)+ (2\delta^{2}-1) sin(3\phi) +6 (2\delta^{2}-3)sin(\phi_{0})\\
- 6 ((2\delta^{2}-1)sin(2 \phi_{0})+2(\phi-\phi_{0}))cos(\phi)
- 2 (2\delta^{2}-1)sin(3 \phi_{0}) )
\end{split}
\end{equation}
\begin{equation}
\begin{split}
I_{2y} = \frac{1}{24} g \sqrt{(1-\delta^{2})}(-3(2\delta^{2}+3)cos(\phi) - (2\delta^{2}-1)cos(3\phi) - 8(2\delta^{2}-1)\\
cos(\phi_{0})^{3} - 6 (2(\phi-\phi_{0})-sin(2 \phi_{0}))sin(\phi) - 12 \delta^{2} (sin(2 \phi_{0}) sin(\phi)))
\end{split}
\end{equation}
%
\begin{equation}
\begin{split}
x(\phi)= \frac{1}{(k p_{z0}-\omega \gamma_{0})}({- p_{x0} \phi - a_{0} \delta (-\phi + sin(\phi))})+ \frac{\tau_{0} a^{2}_{0}}{72} \delta( (9 (-9 + 2 \delta^{2}) cos(\phi)+ (1- 2 \delta^{2})\\ 
 cos(3 \phi) -4 (-20 + 4 \delta^{2} + 9 \phi sin(\phi)))) 
\end{split}
\end{equation}
%
\begin{equation}
\begin{split}
y(\phi)= \frac{1}{(k p_{z0}-\omega \gamma_{0})}({- p_{y0} \phi - a_{0} \delta (-1 + cos(\phi))})+ \frac{\tau_{0} a^{2}_{0}}{72} \sqrt{1-\delta^{2}} (24 (1-\delta^{2}) \phi  + 36 \phi cos(\phi)\\ -9 (7+2\delta^{2})sin(\phi)+(1-2\delta^{2})sin(\phi)) 
\end{split}
\end{equation}
\begin{equation}
\begin{split}
z(\phi)=\frac{1}{13824 (\gamma - p_{z0})}(13824 (p_{z0} \phi + a_{0}(p_{x0} \delta \phi + p_{y0} \sqrt{1-\delta^{2}} (-1+cos(\phi))-p_{x0}\delta sin(\phi)\\
 - \frac{a_{0}(2(1+2\delta^{2})\phi-8 \delta^{2} sin(\phi)+(-1+2\delta^{2})sin(\phi))}{8 (p_{z0}-\gamma_{0})}))+24 \tau_{0} a^{3}_{0} (a_{0}(-99+ 460 \delta^{2} \\ 
 +52 \delta^{4}+72 \phi^{2}-24 (-8 +27 \delta^{2} +2 \delta^{4} )cos(\phi)+96 (-1+2 \delta^{2}) cos(2\phi)+8\delta^{2}(1-2\delta^{2})\\
cos(3\phi)+(1-2\delta^{2})^{2}cos(4\phi)-288\delta^{2}\phi sin(\phi)+72 (-1+2\delta^{2})\phi sin(2\phi))-8(-p_{x0}\delta\\
(9(-9+ 2\delta^{2})cos(\phi)+(1-2\delta^{2})cos(3\phi)-4 (-20+4\delta^{2}+9\phi sin(\phi)))-p_{y0} \sqrt{1-\delta^{2}}\\
(24(1+\delta^{2})\phi  +36 \phi cos(\phi)-9 (7+2\delta^{2})sin(\phi)+(1-2\delta^{2})sin(\phi)+(1-2\delta^{2})
sin(\phi))))\\
-a^{6}_{0} \tau^{2}_{0}(-12 \phi (105+188 \delta^{2}-188 \delta^{4}
 -64 \delta^{6}+24 \phi^{2})+2304(-1+\delta^{4})\phi cos(\phi)
 +1152 \\
 (1-2\delta^{2})\phi cos(2 \phi)-36 (1-2\delta^{2})^{2}\phi cos(4\phi)
-576 (-7 -2 \delta^{2}+7\delta^{4}+ 2 \delta^{6})sin(\phi)\\+9(-1+2\delta^{2})(85 - 20 \delta^{2}  +20 \delta^{4}-48 \phi^{2})sin(2\phi)-64 (1-2\delta^{2} -\delta^{4}+2\delta^{6})sin(3\phi)\\
 +36(1-2\delta^{2})^{2}sin(4\phi) +(-1 + 2 \delta^{2})^{3}sin(6\phi)))
\end{split}  
\end{equation}
%
\bibliography{1stpaper}
\newpage
%
%
\begin{figure}
\centering
\begin{tabular}{cc}
{\includegraphics[width = 3in]{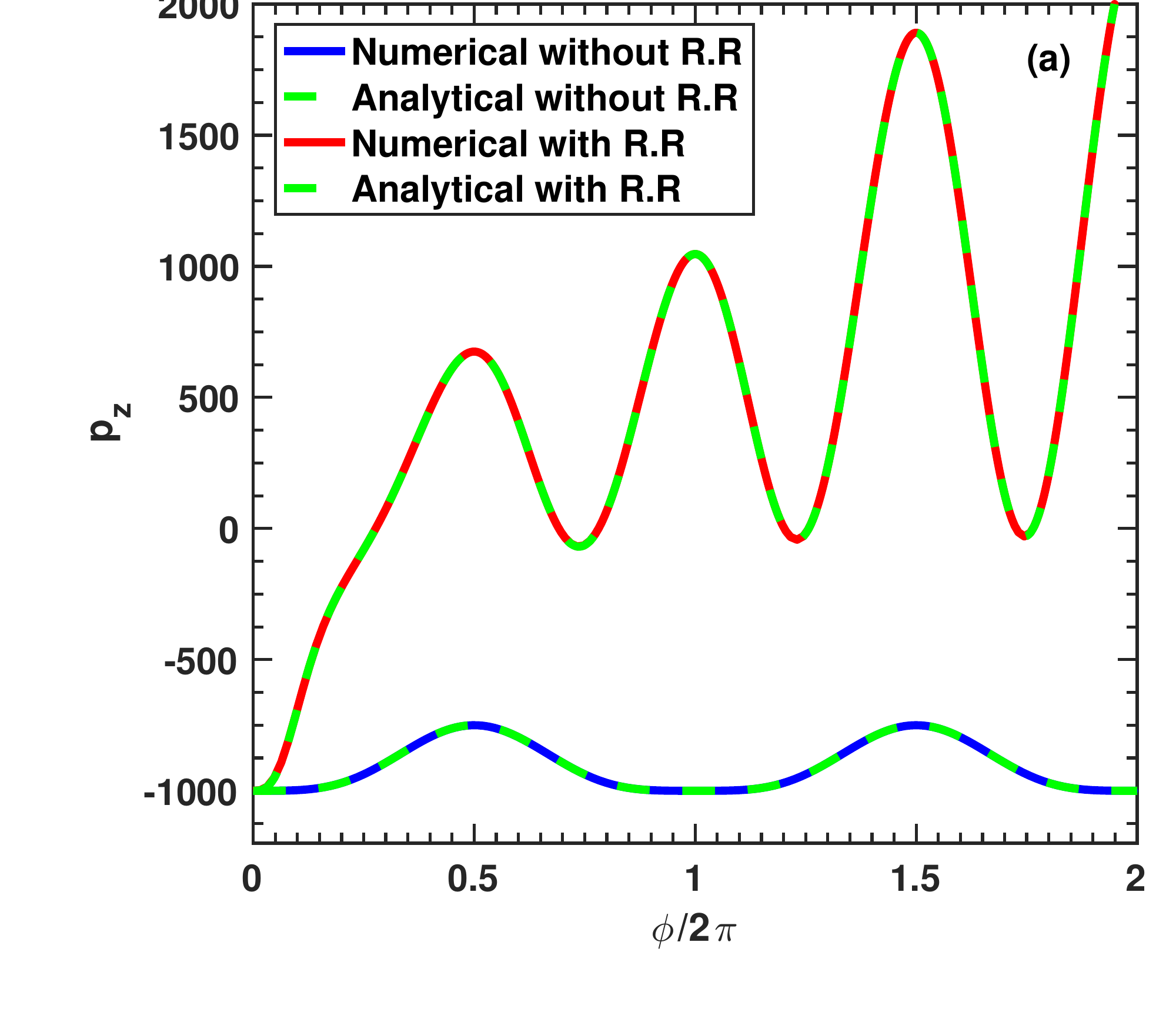}}                & 
{\includegraphics[width = 3in]{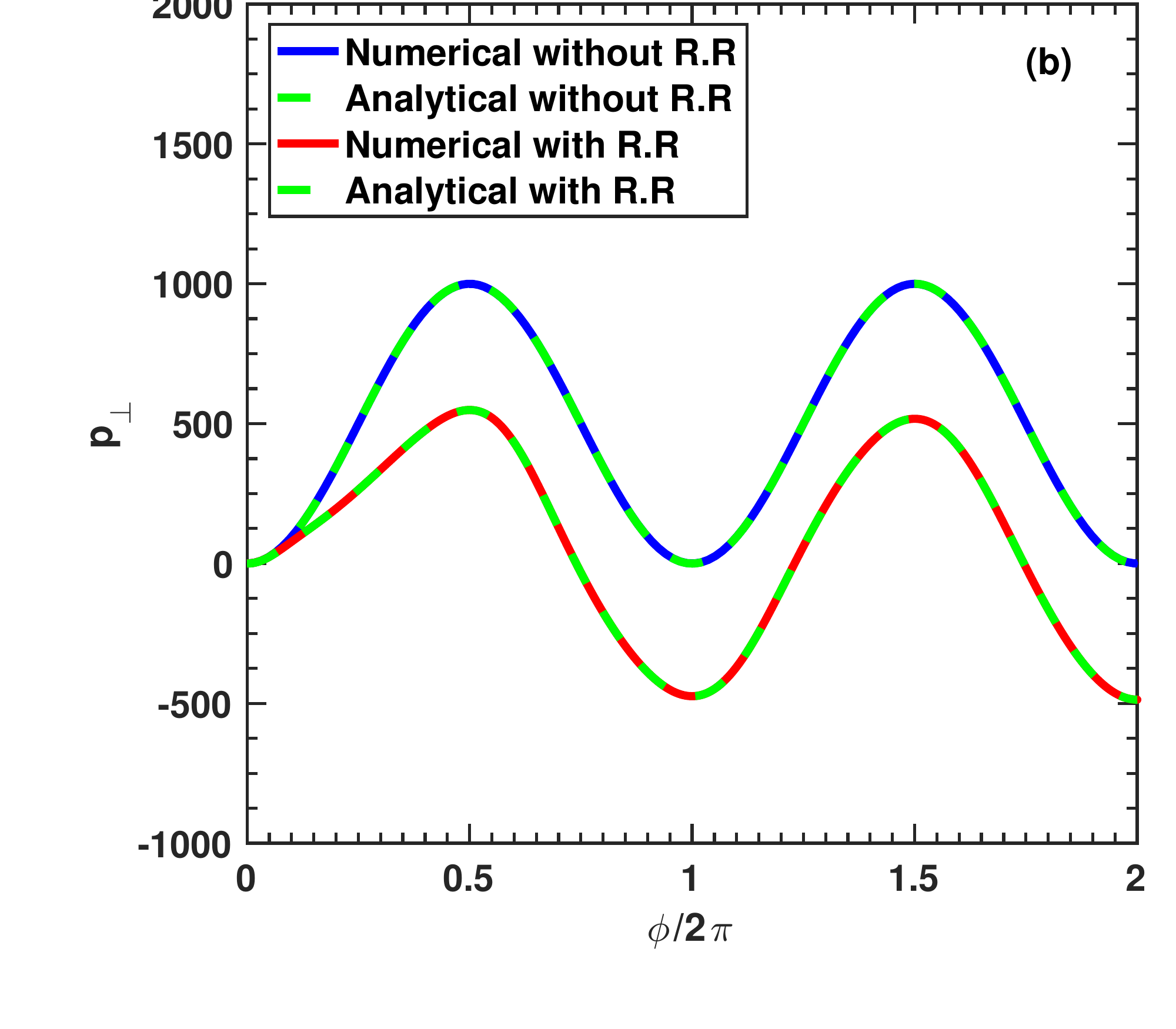}}\\
{\includegraphics[width = 3in]{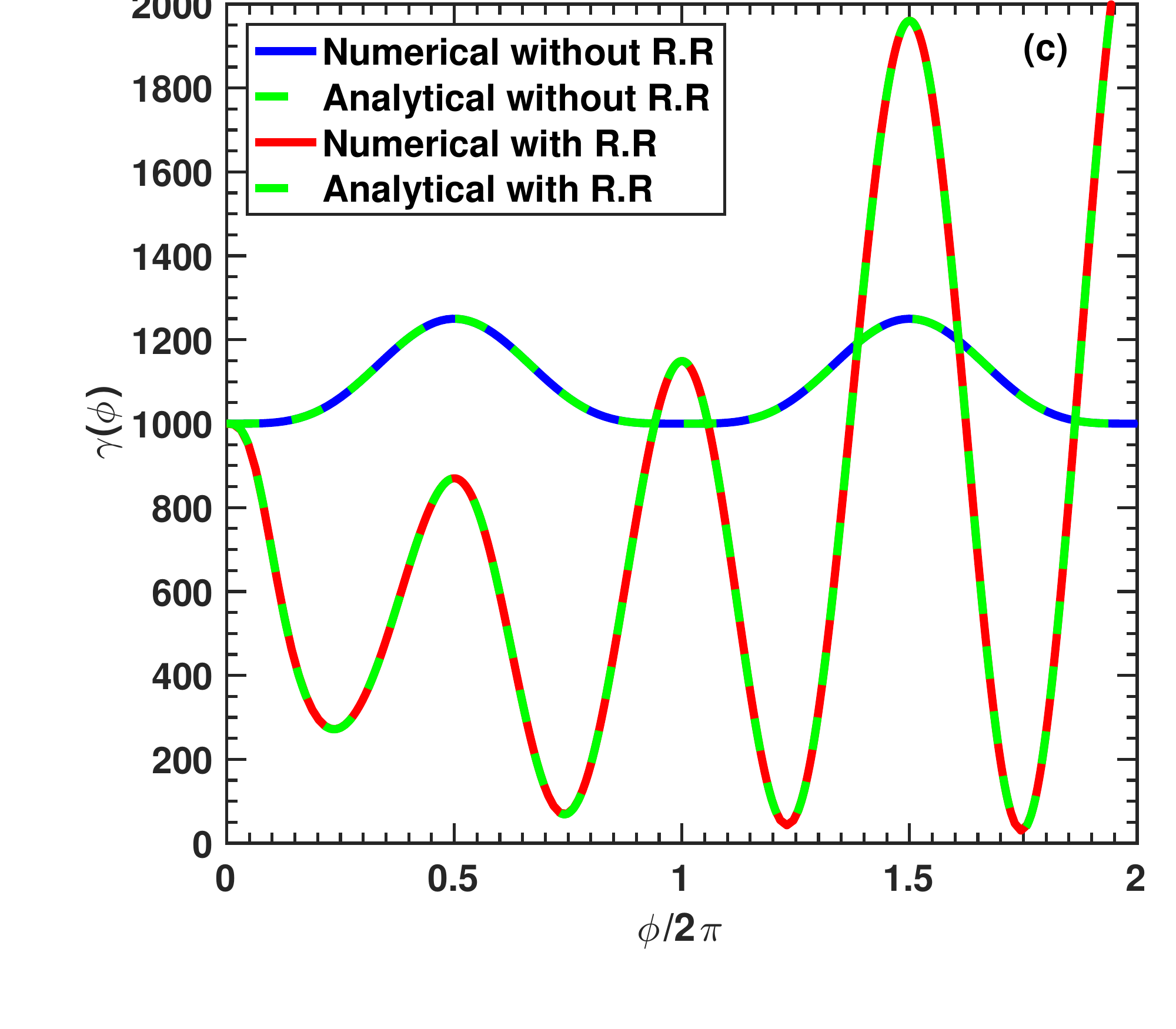}}  &
{\includegraphics[width = 3in]{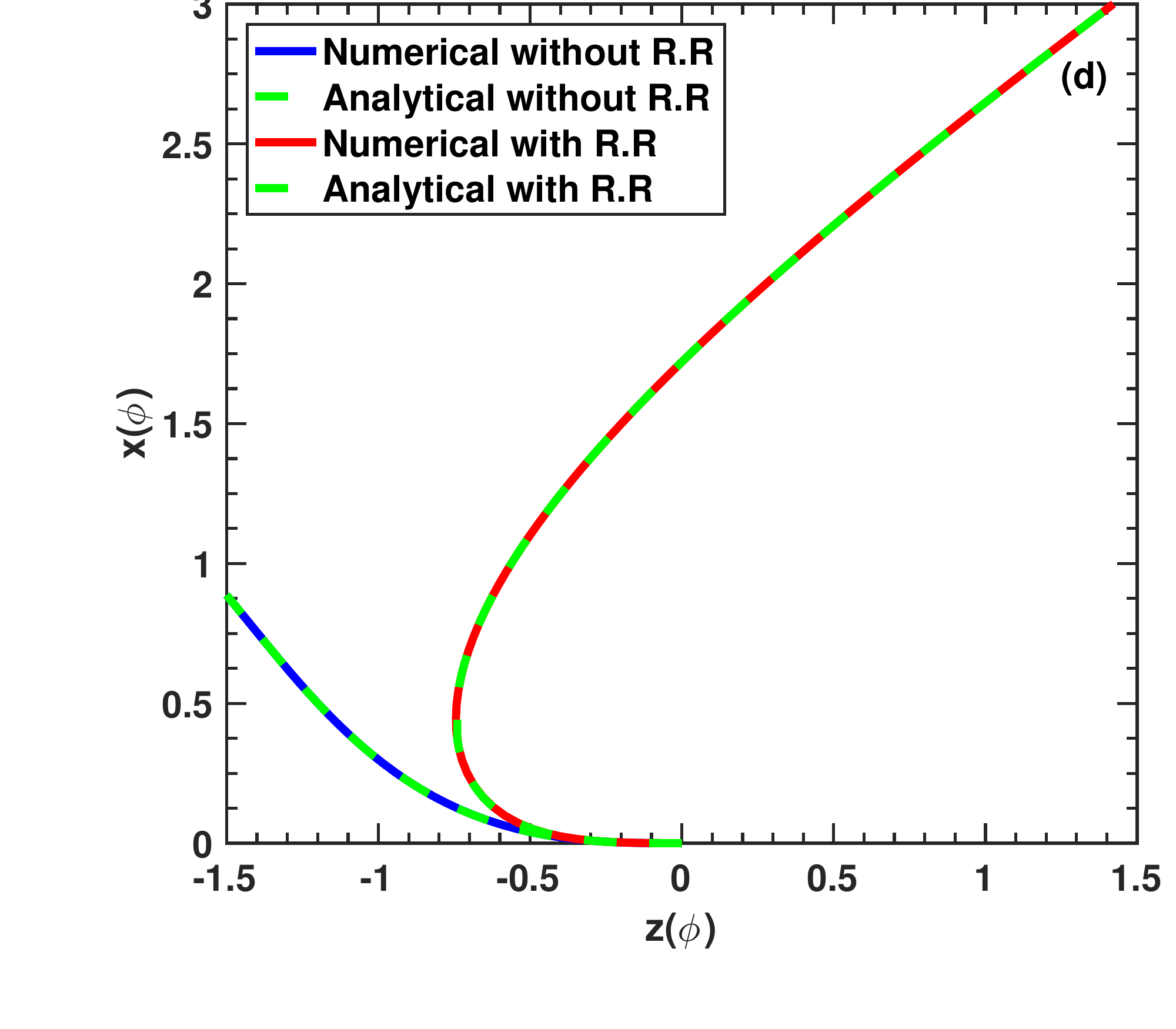}}
\end{tabular}
\caption{Figs. (a), (b), (c) and (d) represent the longitudinal momentum, transverse momentum, energy and trajectory of the particle respectively. The red and blue curves represent the solution of the Hartemann- Luhmann and Lorentz Force equations respectively and the dashed green curves represent the analytical solution. These graphs are plotted for the $a_{0}=500, \delta={1}, p_{z0}=-1000 $.}
\end{figure}
\begin{figure}
\centering
\begin{tabular}{cc}
{\includegraphics[width = 3in]{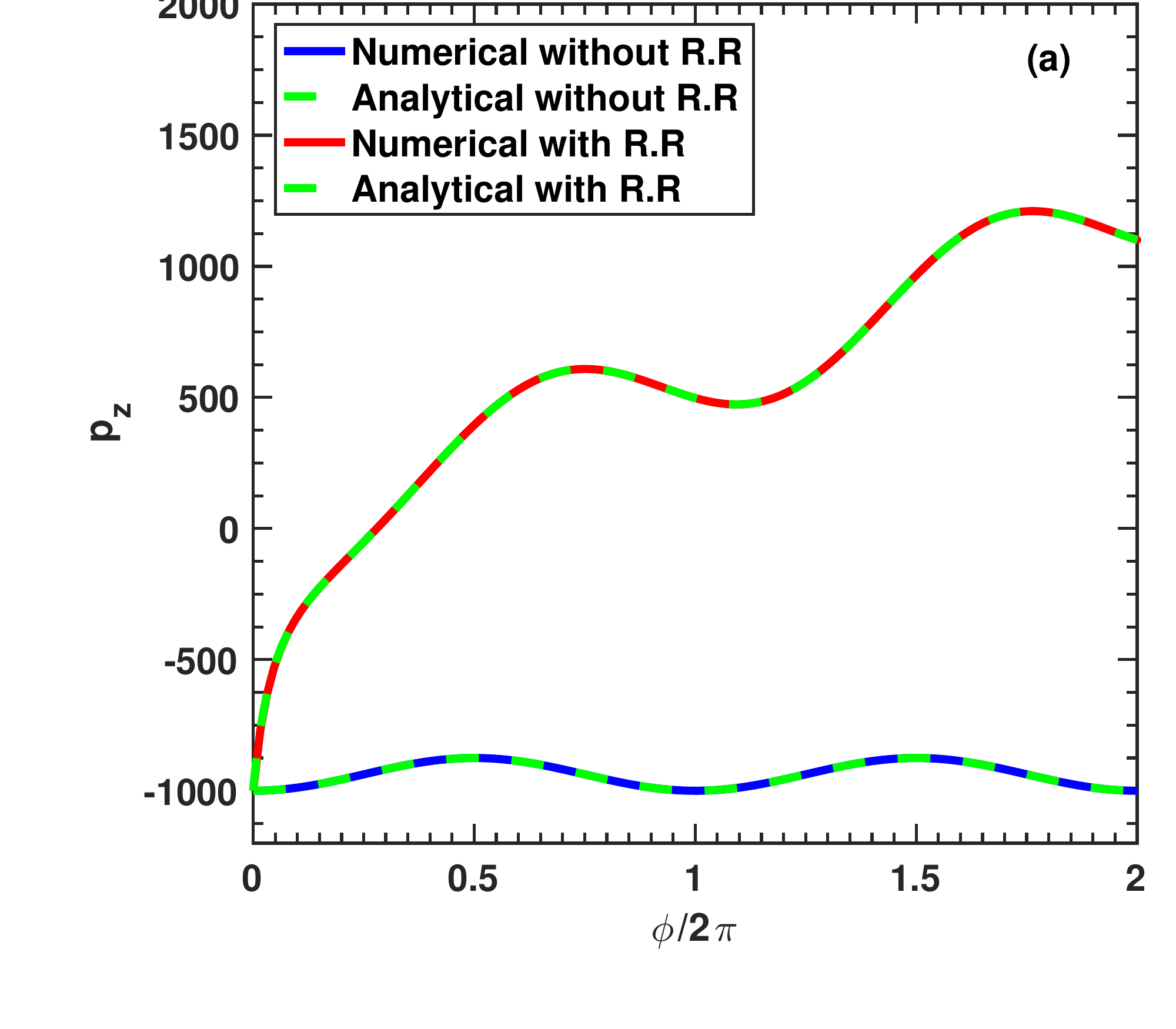}}  &
{\includegraphics[width = 3in]{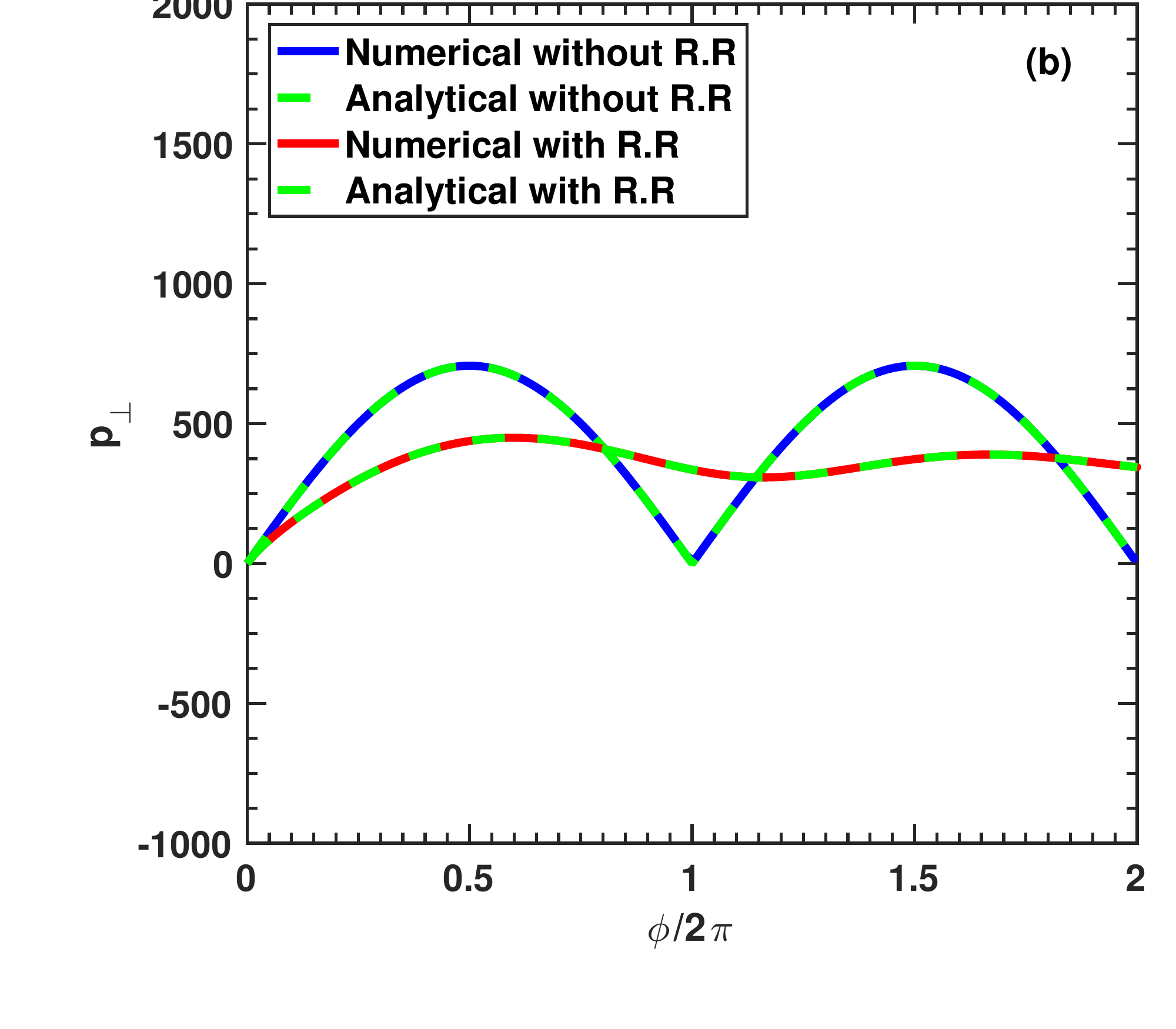}}\\ 
{\includegraphics[width = 3in]{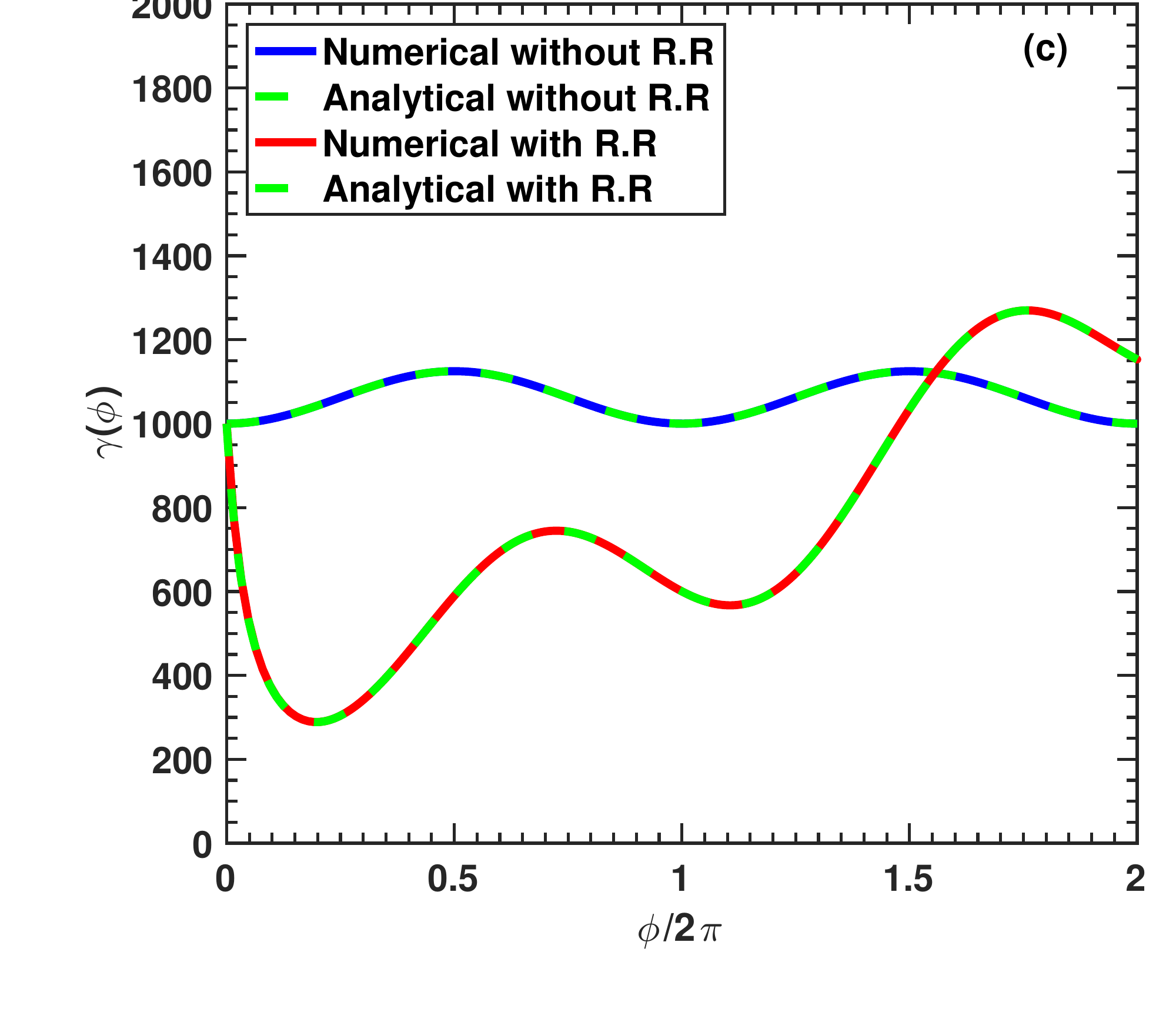}}  &
{\includegraphics[width = 3in]{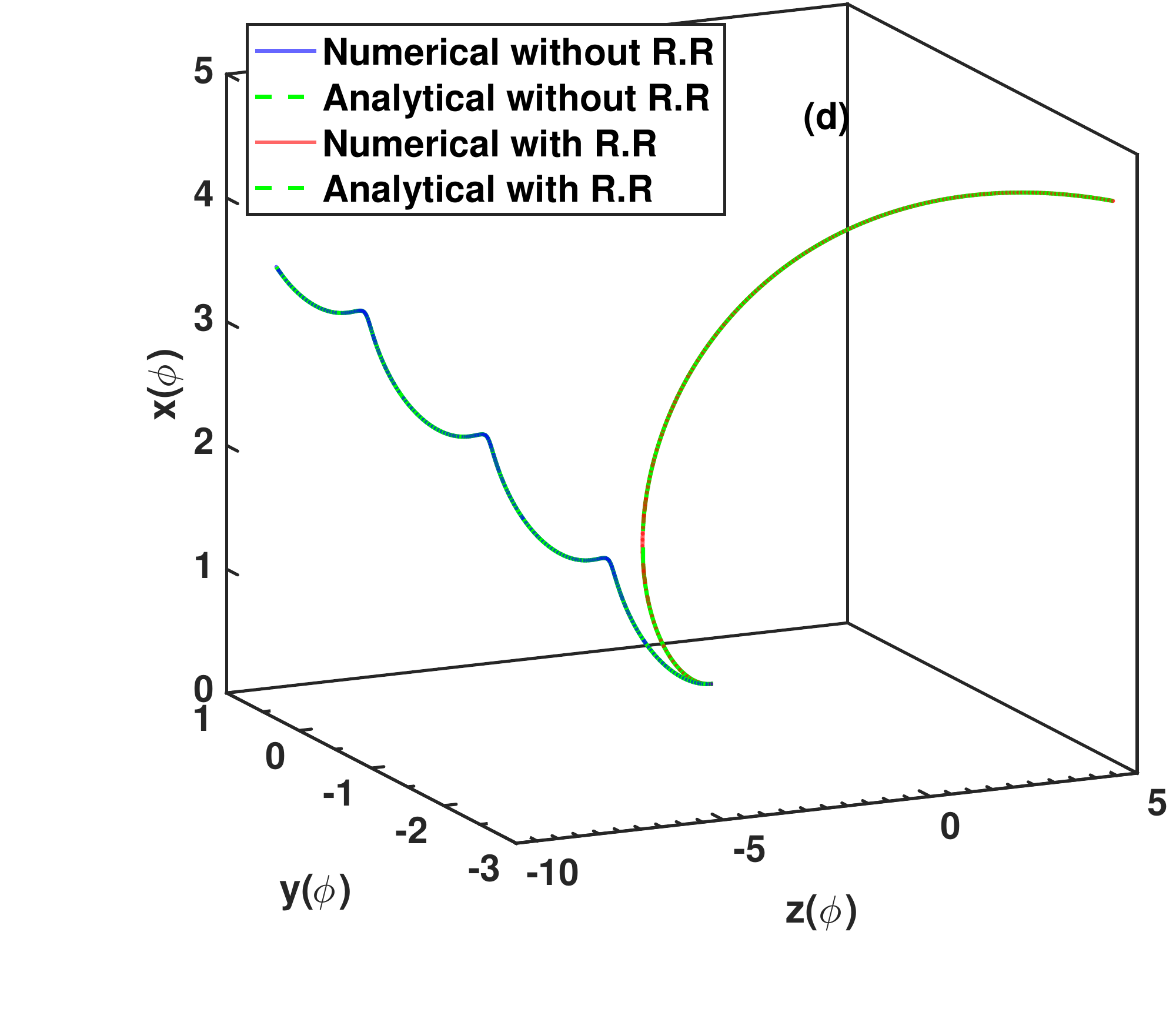}}
\end{tabular}
\caption{Figs. (a),(b),(c) and (d)  represent the longitudinal momentum, transverse momentum, energy and trajectory of the particle respectively. The red and blue curves represent the solution of the Hartemann- Luhmann and Lorentz Force equations respectively and the dashed green curves represent the analytical solution. These graphs are plotted for the $a_{0}=500, \delta=\frac{1}{\sqrt{2}}, p_{z0}=-1000 $. }
\end{figure}
%
\begin{center}
\begin{figure} 
 \includegraphics[width=0.9\linewidth]{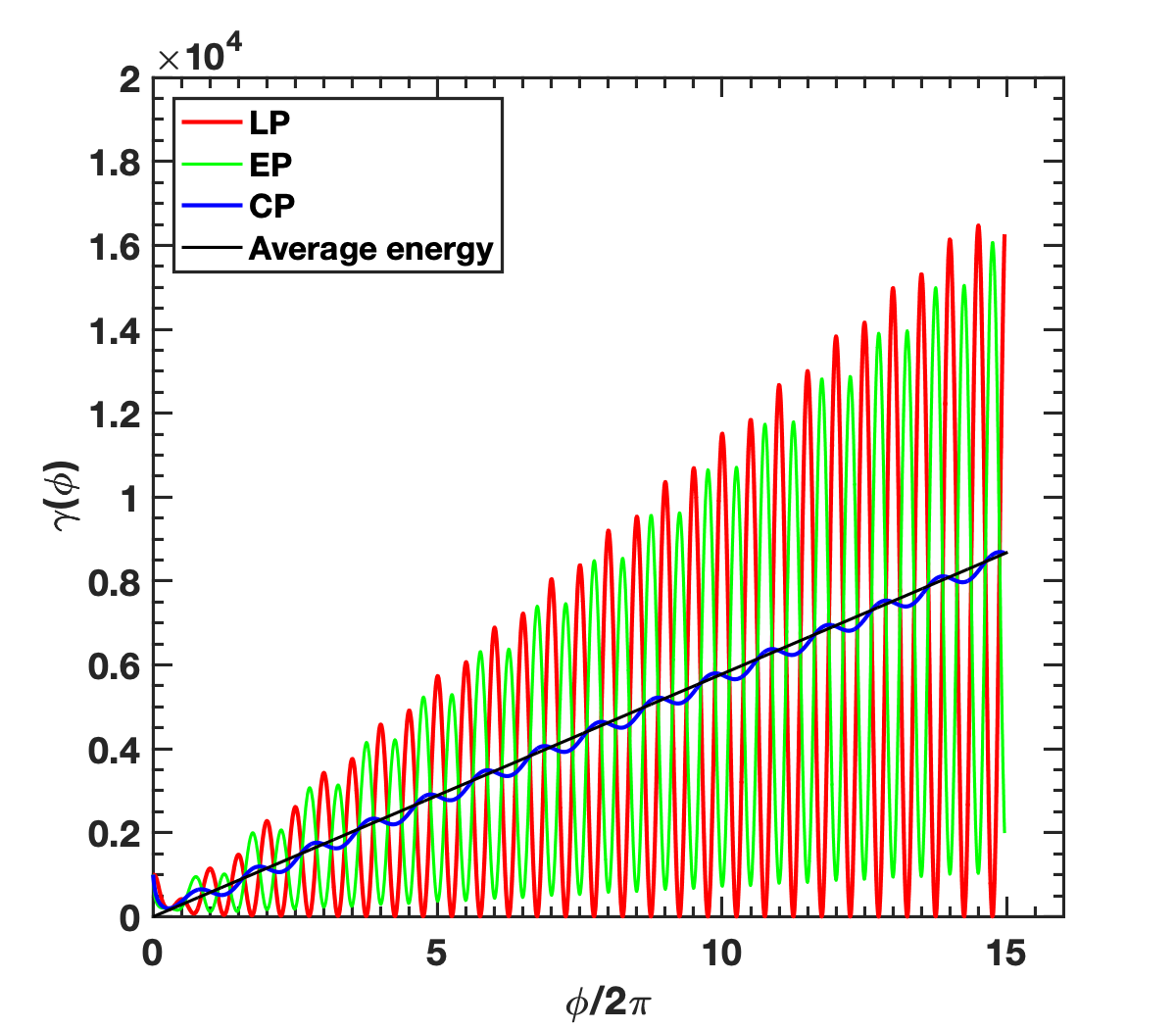}
\caption{Average Energy gain with different polarizations and the red blue and green curves respectively correspond to $\delta = 1,\,\, 1/\sqrt{2},\,\, 1/4$. These graphs are plotted for the $a_{0}=500, p_{z0}=-1000 $. }
\end{figure}
\end{center}

\begin{figure}
\centering
\begin{tabular}{cc}
{\includegraphics[width = 3in]{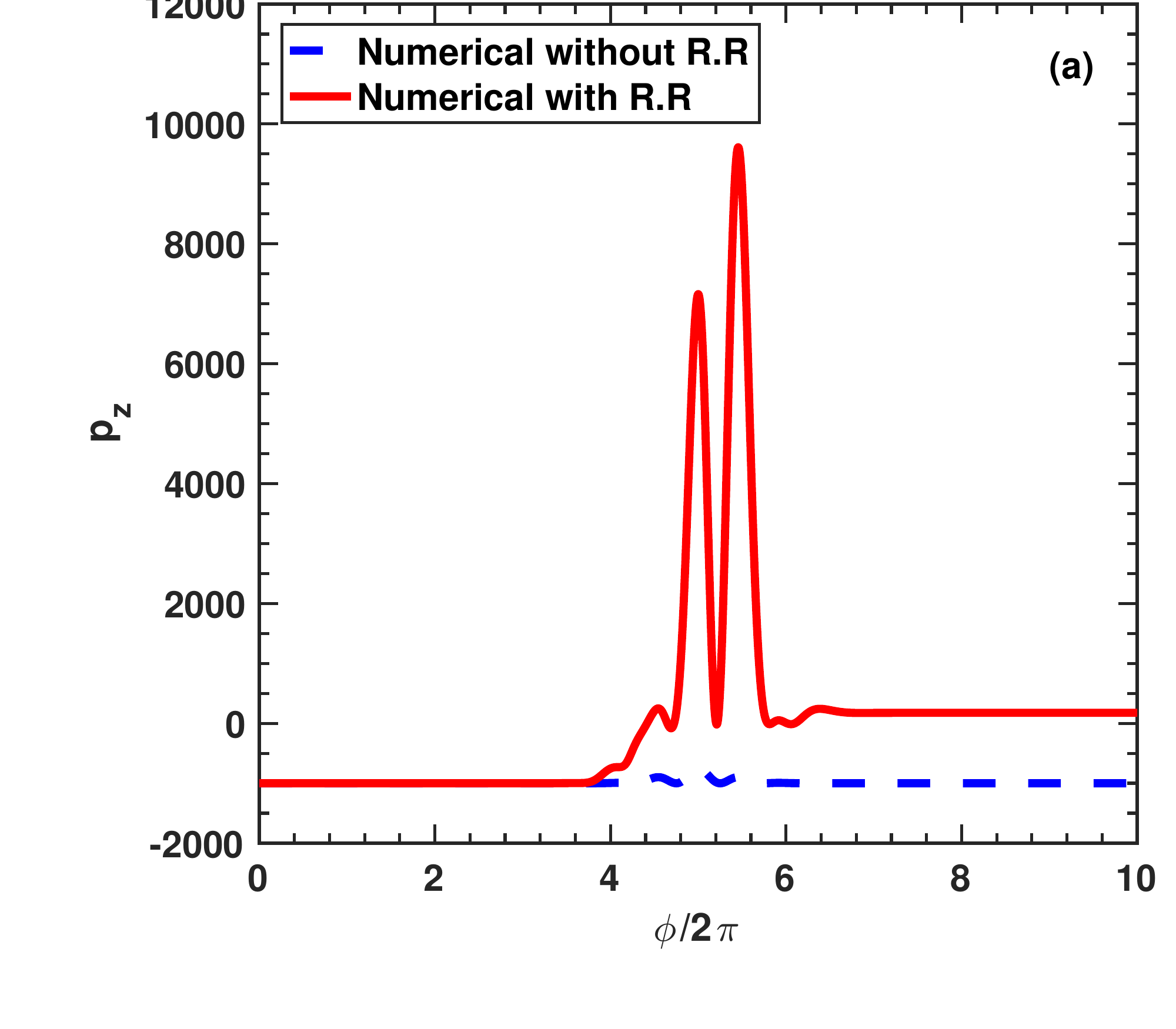}} &
{\includegraphics[width = 3in]{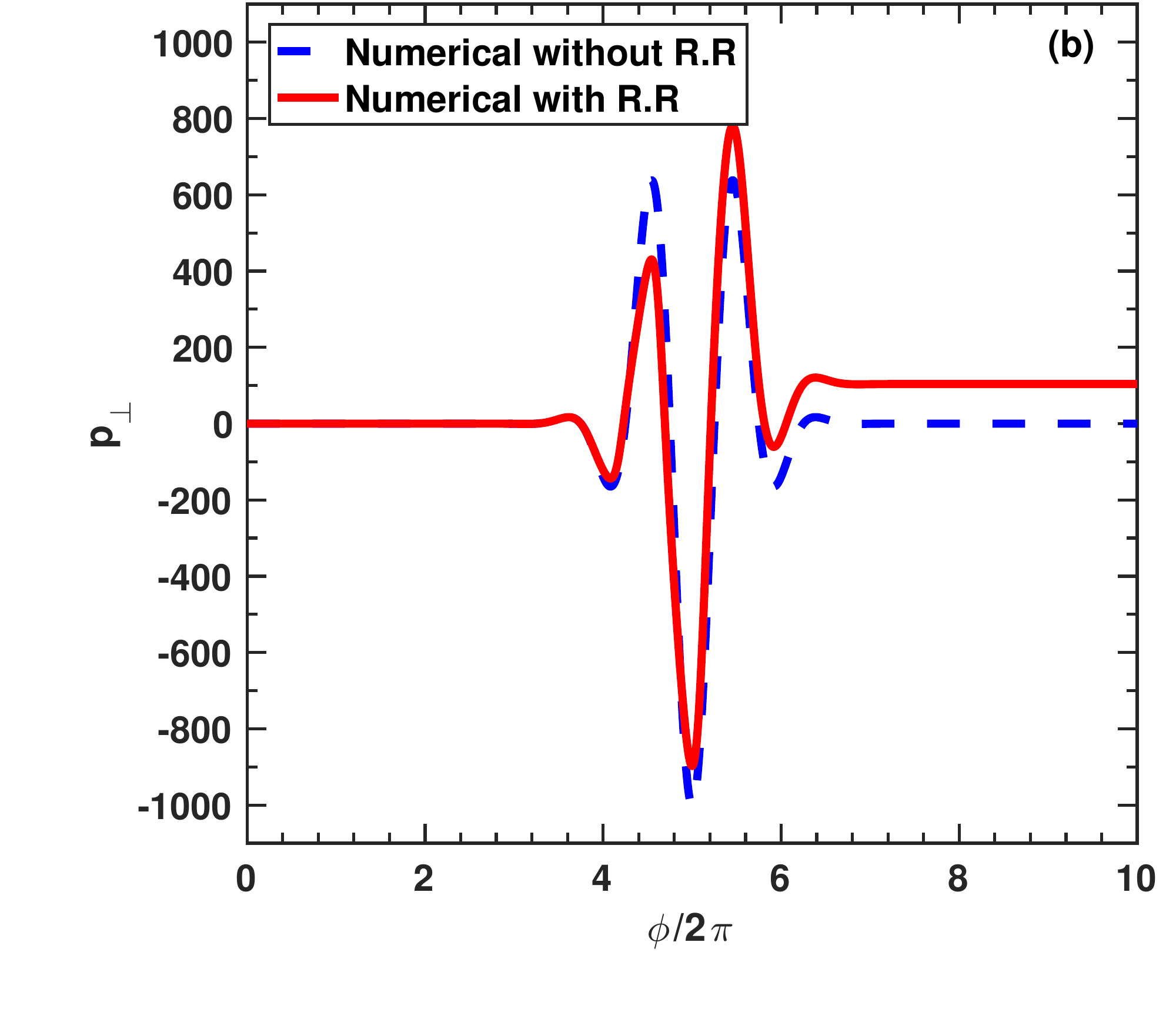}}\\ 
{\includegraphics[width = 3in]{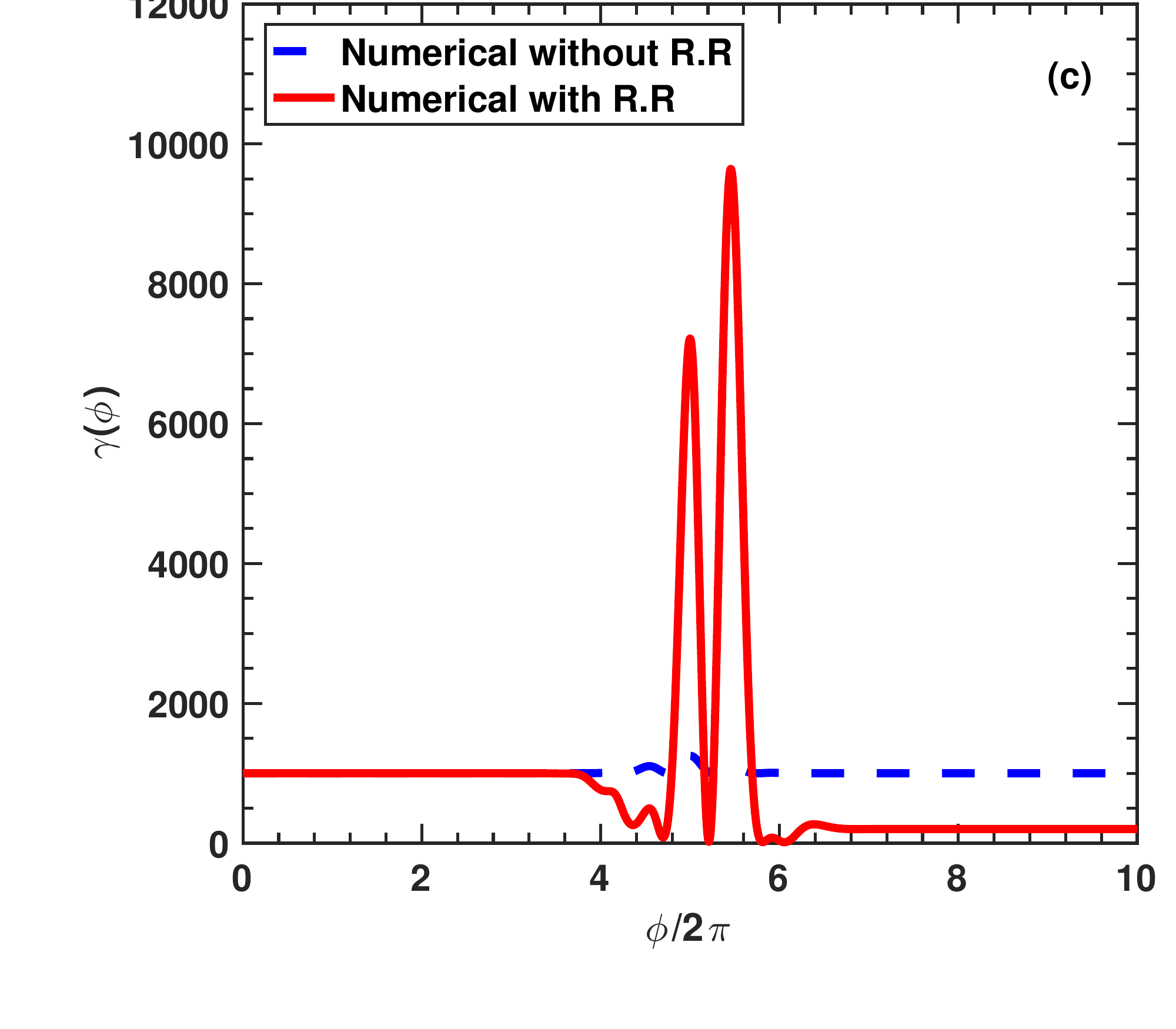}} &
{\includegraphics[width = 2.85in]{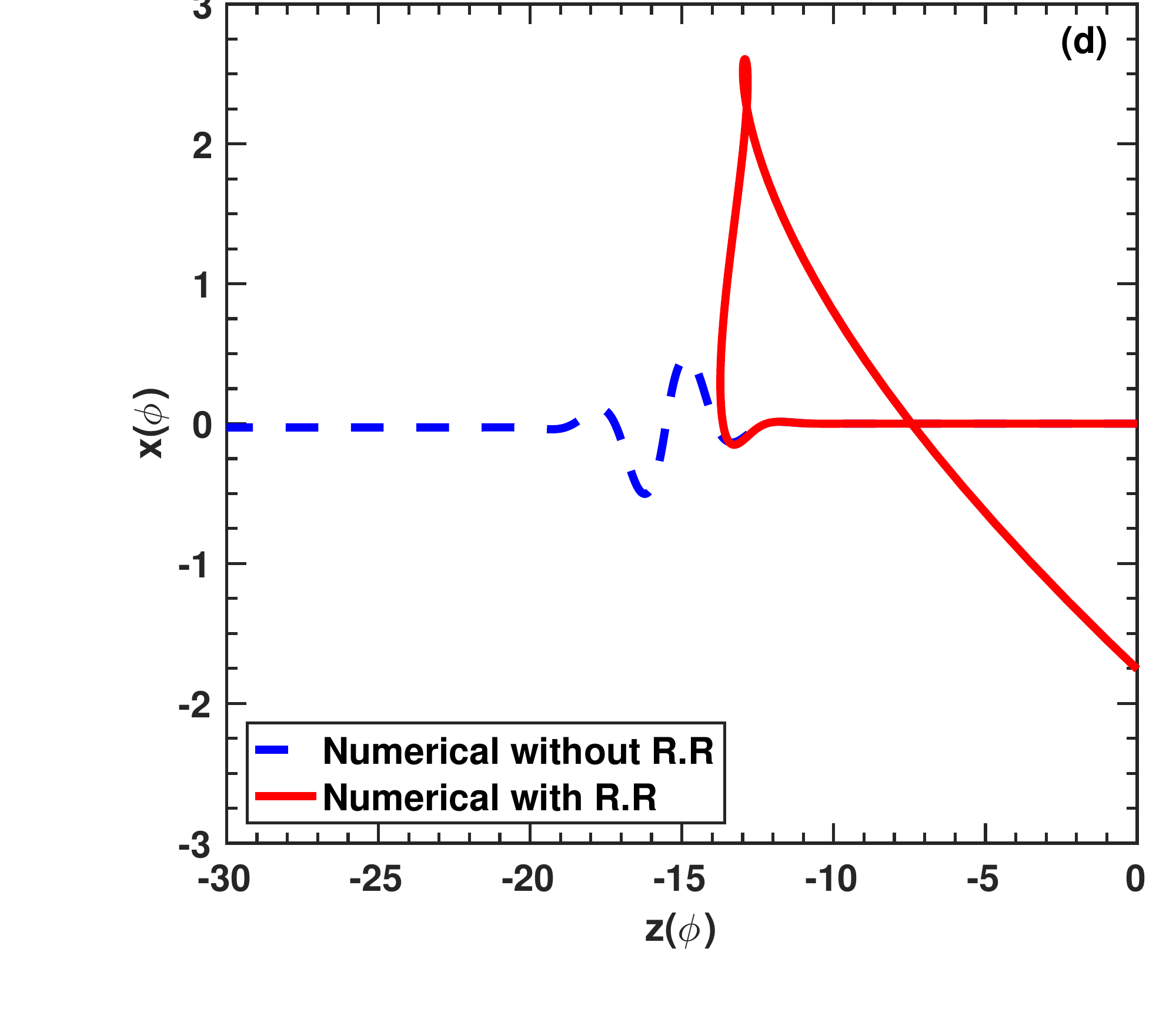}}
\end{tabular}
\caption{Figs. (a), (b), (c) and (d) represent the longitudinal momentum, transverse momentum, energy and trajectory of the particle respectively. The red and blue curves represent the solution of the Hartemann- Luhmann and Lorentz Force equations respectively. These graphs are plotted for the $a_{0}=1000, \delta={1}, p_{z0}=-1000 $. }
\end{figure}
\begin{figure}
\begin{tabular}{cc}
{\includegraphics[width = 3in]{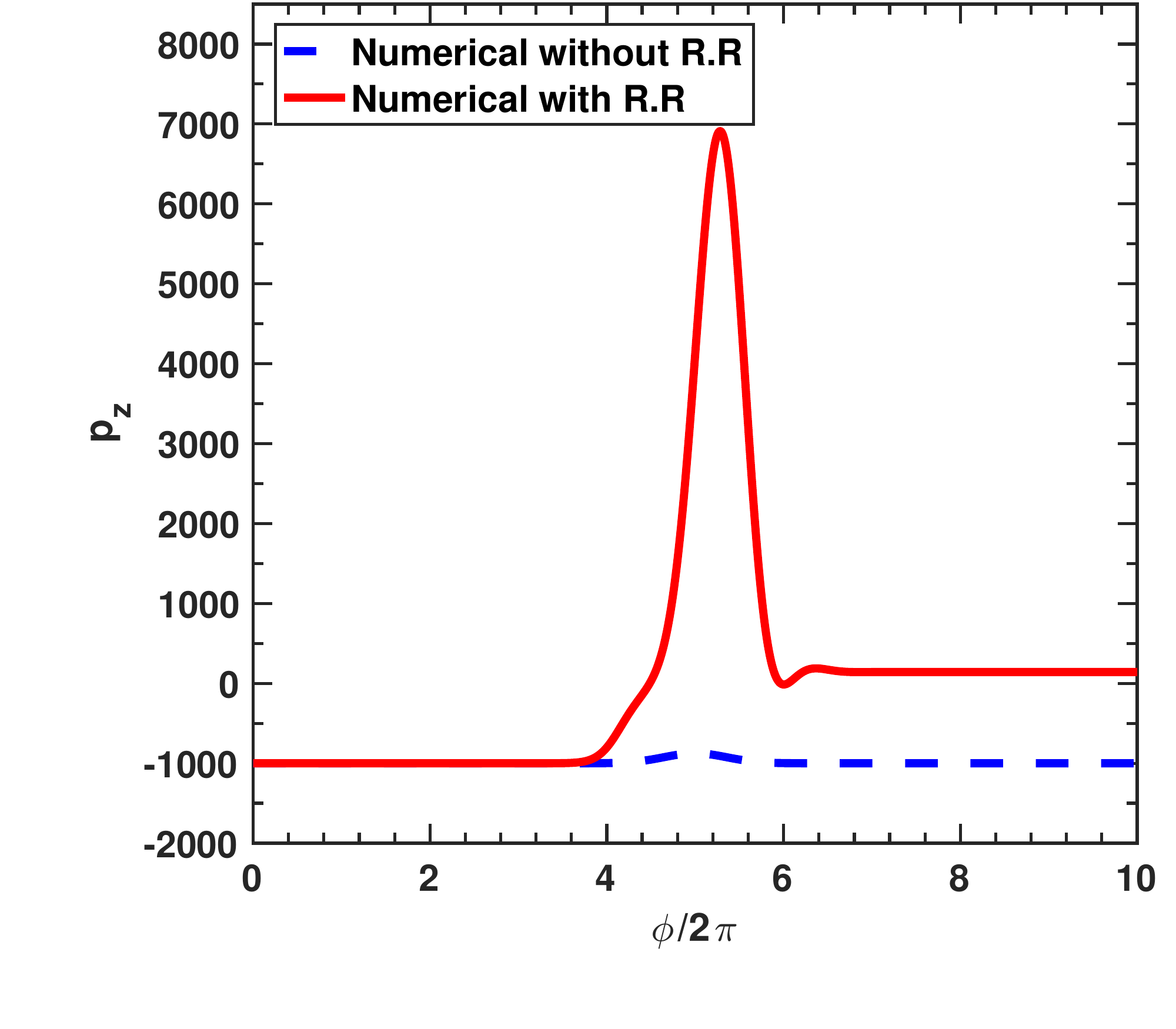}} &
{\includegraphics[width = 3in]{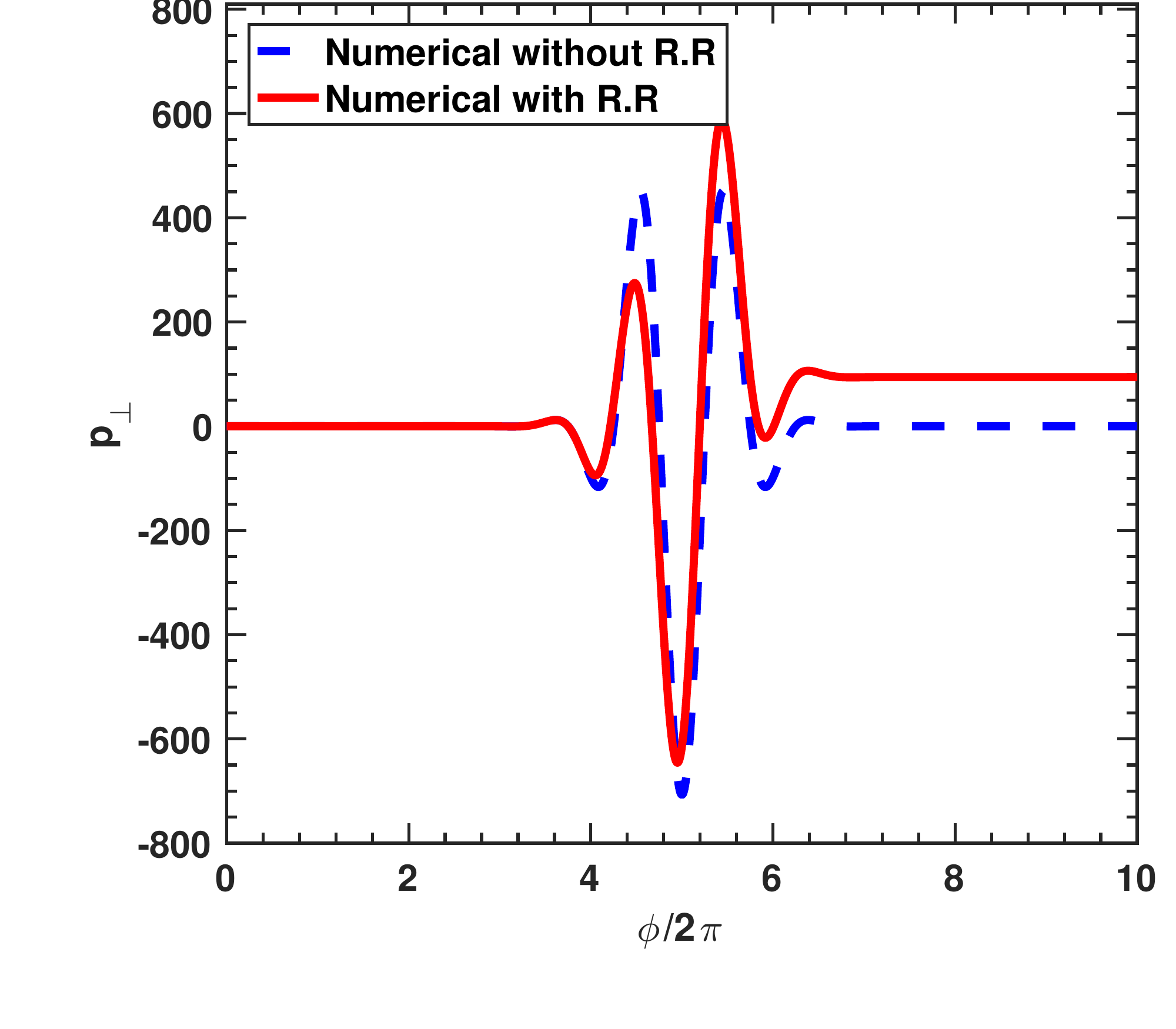}}\\ 
{\includegraphics[width = 3in]{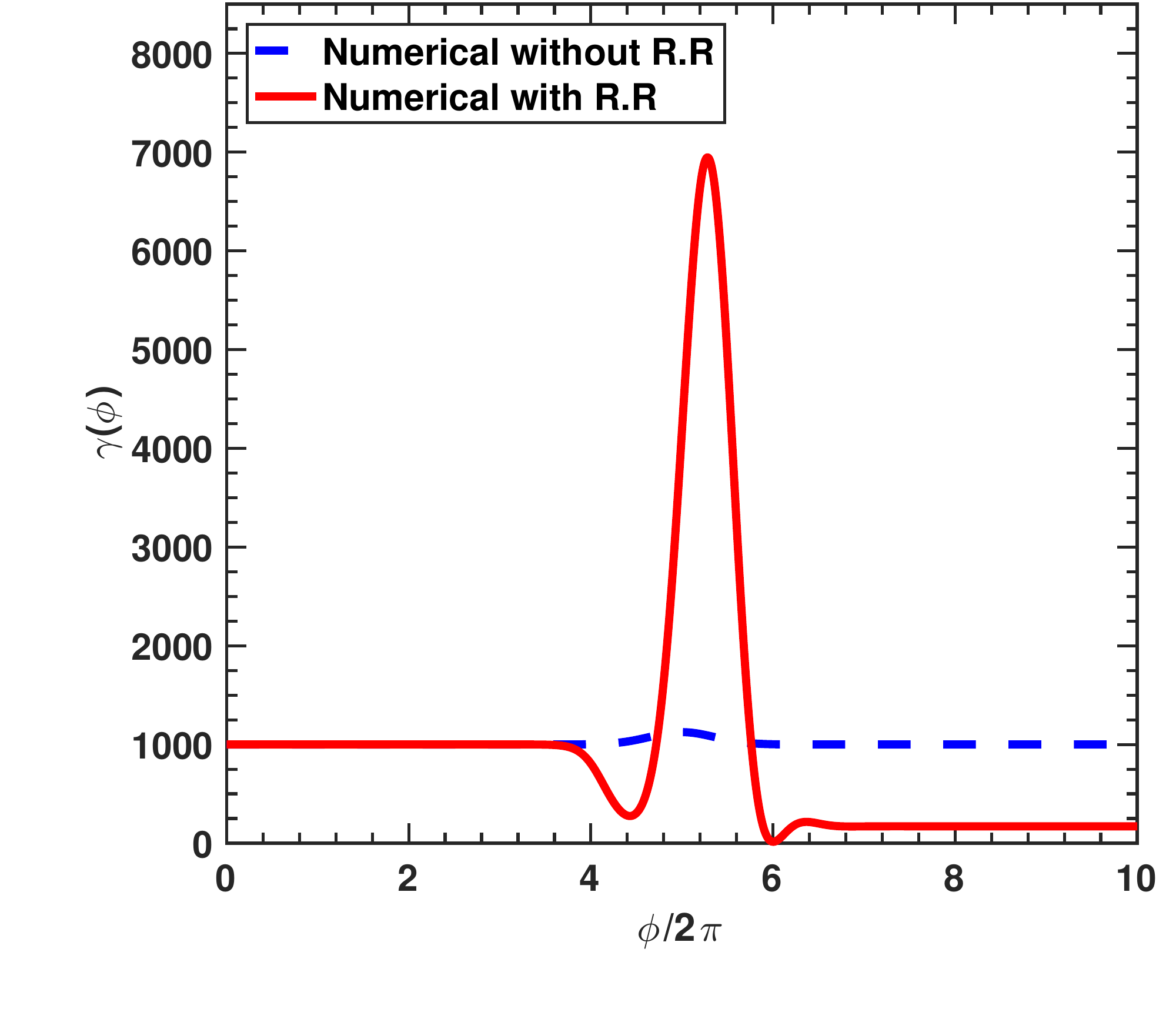}} &
{\includegraphics[width = 3in]{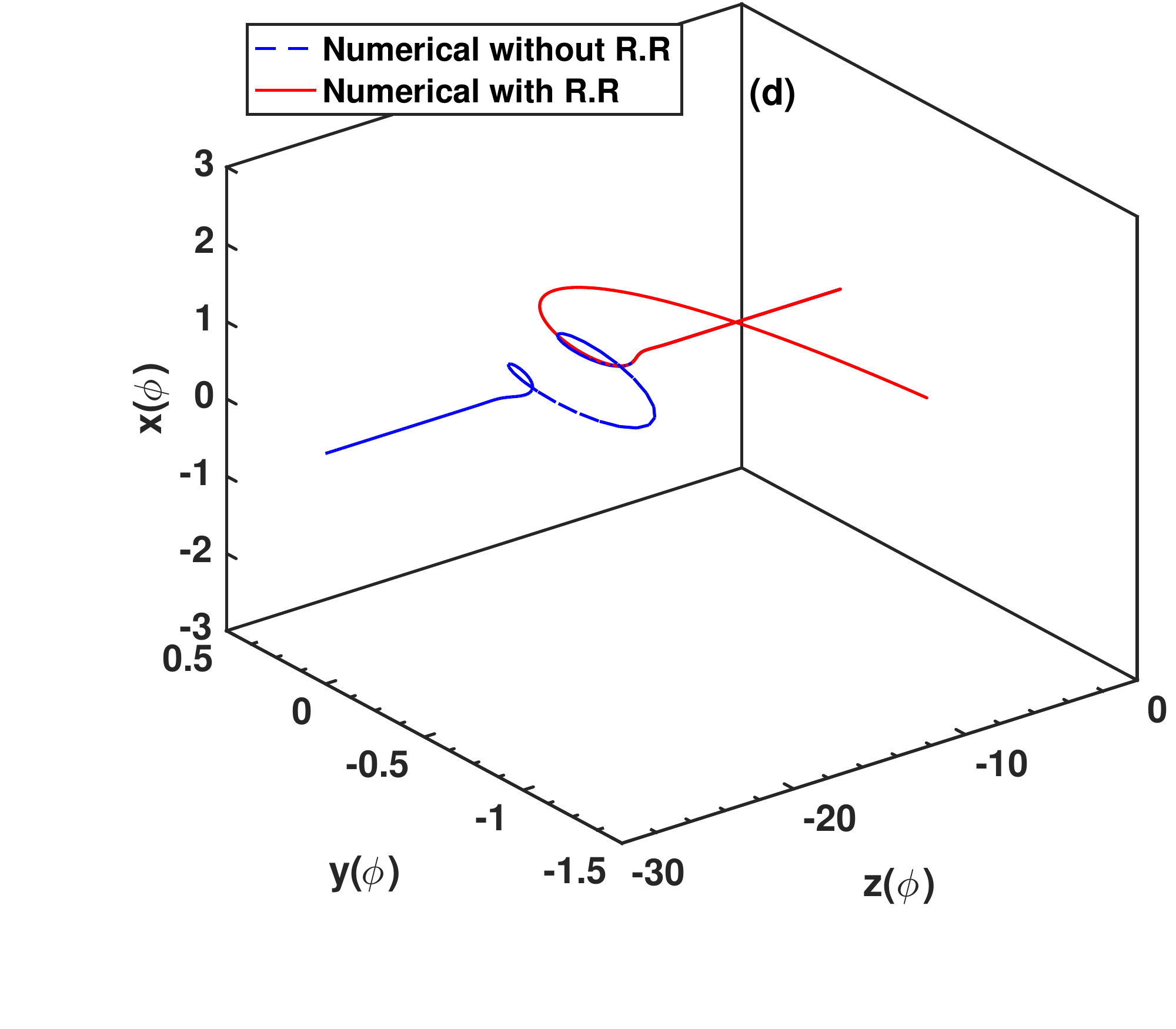}}
\end{tabular}
\caption{Figs. (a), (b), (c) and (d)  represent the longitudinal momentum, transverse momentum, energy and trajectory of the particle respectively. The red and blue curves represent the solution of the Hartemann-Luhmann and Lorentz Force equations respectively. These graphs are plotted for the $a_{0}=1000, \delta=\frac{1}{\sqrt{2}}, p_{z0}=-1000 $. }
\end{figure}
\begin{center}
\begin{figure} 
\includegraphics[width=1\linewidth]{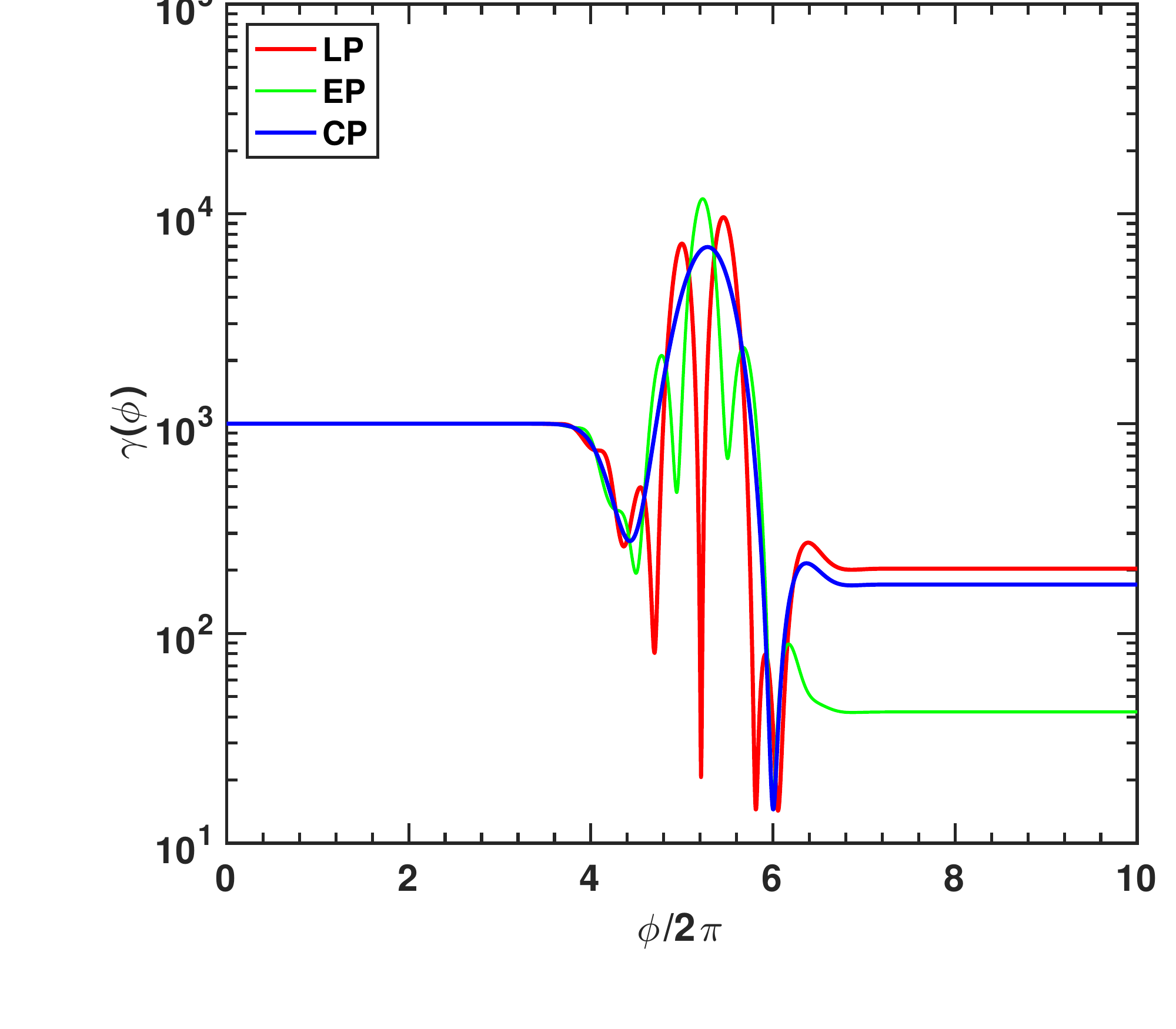}
\caption{Energy gain with different polarizations; the red blue and green curves respectively correspond to $\delta = 1,\,\, 1/\sqrt{2},\,\, 1/4$.  These graphs are plotted for the $a_{0}=1000, p_{z0}=-1000 $.}
\end{figure}
\end{center}
\begin{figure}
\centering
\begin{tabular}{cc}
{\includegraphics[width=0.52\linewidth]{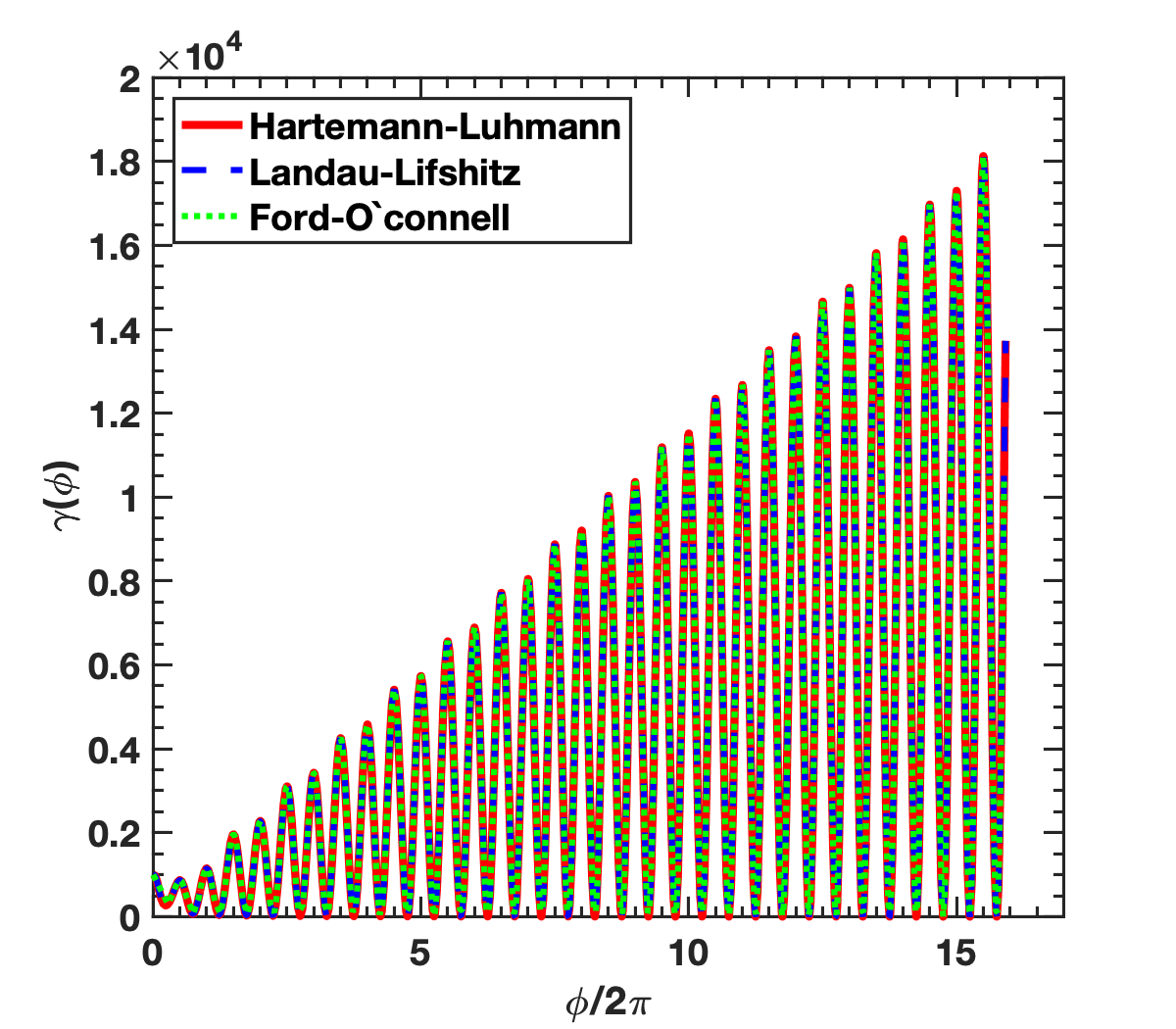}} &
{\includegraphics[width=0.48\linewidth]{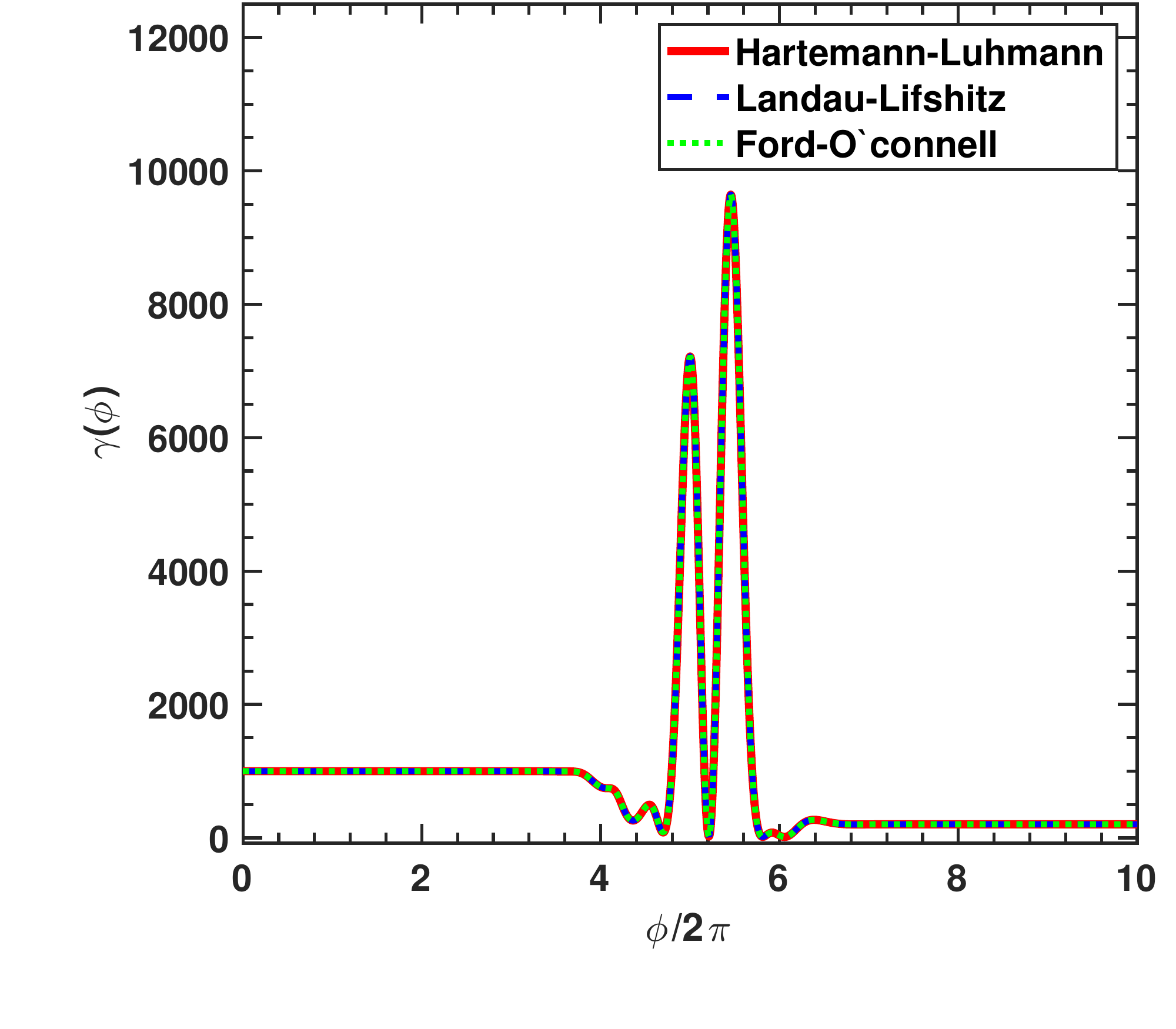}}
\end{tabular}
\caption{Figs. (a), (b)  respectively represent the energy gain computed for a charged particle interacting with a wave train and a Gaussian pulse. The red, blue and green curves respectively represent the solution of the Hartemann-Luhmann, Landau-Lifshitz and Ford-O'connell equations of motion. Here figure7(a) and 7(b) are plotted for the $(a_{0}=500, \delta={1}, p_{z0}=-1000 )$ and $(a_{0}=1000, \delta={1}, p_{z0}=-1000 )$ respectively.}
\end{figure} 
\end{document}